\documentclass[11pt]{preprint} 
\usepackage{amsmath}
\usepackage{amsfonts}
\usepackage{amssymb}

\usepackage{times}
\usepackage{mhfig}
\usepackage{Symbols}

\DeclareMathOperator{\Div}{\mathsf{div}}
\DeclareMathOperator{\Curl}{\mathsf{curl}}

\newcommand{\ZZ}{\mathbb{Z}}

\newtheorem{theorem}{Theorem}[section]

\newtheorem{condition}{Condition}

\newtheorem{corollary}[theorem]{Corollary}

\newtheorem{definition}[theorem]{Definition}
\newtheorem{example}[theorem]{Example}

\newtheorem{lemma}[theorem]{Lemma}

\newtheorem{proposition}[theorem]{Proposition}
\newtheorem{remark}[theorem]{Remark}

\newenvironment{proof}[1][Proof]{\textbf{#1.} }{\ \rule{0.5em}{0.5em}}

\begin{document}


\date{} \title{Remarks on the K41 scaling law in turbulent fluids}
\author{F. Flandoli\inst{1}, M. Gubinelli\inst{1}, M. Hairer\inst{2}, M. Romito\inst{3}}
\institute{Department of Applied Mathematics, University of Pisa
\and Department of Mathematics, The University of Warwick, Coventry CV4 7AL, United
Kingdom.
\and Department of Mathematics, University of Florence}
\titleindent=0.65cm

\maketitle \thispagestyle{empty}

\begin{abstract}
A definition of K41 scaling law for suitable families of measures is given and
investigated. First, a number of necessary conditions are proved. They imply
the absence of scaling laws for 2D stochastic Navier-Stokes equations and for
the stochastic Stokes (linear) problem in any dimension, while they imply a
lower bound on the mean vortex stretching in 3D. Second, for 3D stochastic
Navier-Stokes equations necessary and sufficient conditions for K41 are proved,
translating the problem into bounds for energy and enstrophy of high and low
modes respectively. The validity of such conditions in 3D remains open.
Finally, a stochastic vortex model with such properties is presented.
\end{abstract}

%

%

\section{Introduction}

In very rough terms, the scaling law devised by Kolmogorov and Obukhov for
turbulent 3D fluids (usually referred as K41, see~\cite{Kol} and a detailed
discussion in~\cite{Frisch}), says that $S_{2}\left(  r\right)  \sim\epsilon
^{2/3}r^{2/3}$ where $S_{2}\left(  r\right)  $ is the second order structure
function and $\epsilon$ is the mean energy dissipation rate. Moreover, it is
specified that this law is valid at very high Reynolds numbers and for
distances $r$ in a certain range between the integral range and the Kolmogorov
dissipation scale, having the order $\eta=\nu^{3/4}\epsilon^{-1/4} $. Although
the numerical essence of these claims may be clear, their precise mathematical
interpretation is not necessarily unique and could change a little bit
depending on new discoveries.

The purpose of this note is to give one possible precise mathematical
formulation of this scaling law and to discuss it from a number of viewpoints. We
immediately stress that we cannot prove its validity for the 3D Navier-Stokes
equations, but nevertheless we obtain a number of insights that seem worth
to be known.

Some of our considerations are true for quite general families of probability
measures; others will be specific to the stochastic Navier-Stokes equations on
the torus $\left[  0,1\right]  ^{d}$, $d=2,3$,
\begin{equation}
\frac{\partial u}{\partial t}+\left(  u\cdot\nabla\right)  u+\nabla
p=\nu\triangle u+\sum_{\alpha} h_{\alpha}(x) {\dot\beta}_{\alpha}(t)
 \label{stochNS}%
\end{equation}
with $\Div u=0$ and periodic boundary conditions, with suitable vector fields $h_{\alpha}\left(  x\right)  $ and independent Brownian motions $\beta_{\alpha}\left(
t\right)  $ (the torus instead of a more realistic framework has been chosen
for mathematical simplicity). We consider this equation in the limit $\nu\rightarrow
0$. Since the force does not vanish as $\nu\rightarrow0$, this is a singular
limit problem much like the boundary layer one, and so may be considered as a
prototype of high Reynold number singular limit problem, with some
mathematical simplification due to the advantages produced by stochastic analysis. It should be noted that another possible and interesting approach to the zero-viscosity limit is the one adopted in~\cite{Kuksin} (for the $2d$ case), where the amplitude of the forcing noise is proportional to the square-root of the viscosity.

In the following we shall use the parameter $\nu^{-1}$ in place of the
Reynolds number; this simplification is justified in our model since the force
and the domain are given, so the Reynolds number goes to infinity if and only
if $\nu\rightarrow0$.

We shall denote by $H$ the natural space of finite energy velocity fields on
the torus and we shall introduce a space $\mathcal{P}$ of probability measures
on $H$ having certain symmetries and regularities (precise definitions are
given in the next section). On the fields $\varphi^{\left(  i\right)  }\left(
x\right)  $ we shall assume conditions such that there exists at least one
stationary probability measure $\mu\in\mathcal{P}$ associated to
(\ref{stochNS}) (stationary measures will be defined in the next section). We
use the notation
\[
\mu\left[  f\left(  u\right)  \right]  :=\int_{H}f\left(  u\right)
d\mu\left(  u\right)
\]
whenever the integral is well defined.

For every $\mu\in\mathcal{P}$ we introduce the \textit{second order structure
function}
\begin{equation}
S_{2}^{\mu}\left(  r\right)  =\mu\left[  \left\|  u\left(  r\cdot e\right)
-u\left(  0\right)  \right\|  ^{2}\right] \label{struct}%
\end{equation}
for some coordinate unitary vector $e$, with $r>0$ (the results proved below
extend to the so called longitudinal structure function; we consider
(\ref{struct}) to fix the ideas). The measures of $\mathcal{P}$ are supported
on continuous vector fields, so the pointwise operations in (\ref{struct}) are
meaningful. Moreover, the symmetries in
$\mathcal{P}$\ imply that $S_{2}^{\mu}\left(  r\right)  $ is independent of
the coordinate unitary vector $e$ (in addition most of the estimates proved in
the sequel extend to every unitary vector $e$).

We are going to define K41 scaling law for a set $\mathcal{M}\subset
\mathcal{P}\times\mathbb{R}_{+}$. The reason is that equation (\ref{stochNS})
may have (a priori) more than one stationary measure for any given $\nu$ and
in certain claims it seems easier to consider a set of measures for a given
$\nu$. Given $\nu>0$ we use the notation $\mathcal{M}_{\nu}$ for the set
section $\left\{  \mu\in\mathcal{P}:\left(  \mu,\nu\right)  \in\mathcal{M}%
\right\}  $.

Given $\left(  \mu,\nu\right)  \in\mathcal{P}\times\mathbb{R}_{+}$, we define
the \textit{mean energy dissipation rate} as
\[
\epsilon=\epsilon\left(  \mu,\nu\right)  :=\nu\cdot\mu\left[  \int_{\left[
0,1\right]  ^{d}}\left\|  Du\left(  x\right)  \right\|  ^{2}dx\right]  .
\]

\begin{remark}
If $\mu$ is a stationary measure of (\ref{stochNS}) and a mean energy
\textit{equality} (coming from It\^{o} formula)\ can be rigorously proved, one
can show that $\epsilon$ does not depend on $\left(  \mu,\nu\right)  $.
\end{remark}

\bigskip Given $\left(  \mu,\nu\right)  \in\mathcal{P}\times\mathbb{R}_{+}$,
we also define the quantity
\[
\eta=\eta\left(  \mu,\nu\right)  :=\nu^{3/4}\epsilon\left(  \mu,\nu\right)
^{-1/4}.
\]

\begin{remark}
In case of equations (\ref{stochNS}), $\eta$ is a length scale: $\nu$ has
dimension $\left[  L\right]  ^{2}\left[  T\right]  ^{-1}$, $\epsilon$ has
dimension $\left[  L\right]  ^{2}\left[  T\right]  ^{-3}$, so $\eta$ has
dimension $\left[  L\right]  $. The only combination of $\nu$ and $\epsilon$
in powers, having dimension $\left[  L\right]  $, is the $\eta$ above. This is
the simplest reason to choose $\eta$ as a length scale involved in K41 theory.
More refined arguments may be found in~\cite{Frisch} and related references.
\end{remark}

Let us come to the definition of K41 scaling law chosen in this work. Here and
in the sequel, when we talk about a set $\mathcal{M}\subset\mathcal{P}%
\times\mathbb{R}_{+}$, we tacitly assume that
\[
\mathcal{M}_{\nu}\neq\emptyset\text{ for all sufficiently small }\nu>0\;,
\]
since otherwise several definitions and statements would be just empty.

\begin{definition}
\label{defk41}We say that a scaling law of K41 type holds true for a set
$\mathcal{M}\subset\mathcal{P}\times\mathbb{R}_{+}$ if there exist $\nu_{0}%
>0$, $C>c>0$, $C_{0}>0$, and a monotone function $R_{0}:(0,\nu_{0}%
]\rightarrow\mathbb{R}_{+}$ with $R_{0}\left(  \nu\right)  >C_{0}$ and
$\lim_{\nu\rightarrow0}R_{0}\left(  \nu\right)  =+\infty$, such that the bound
\[
c\cdot r^{2/3}\leq S_{2}^{\mu}\left(  r\right)  \leq C\cdot r^{2/3}\text{ }%
\]
holds for every pair $\left(  \mu,\nu\right)  \in\mathcal{M}$ and every $r$
such that $\nu\in(0,\nu_{0}]$ and
\[
C_{0}\cdot\eta\left(  \mu,\nu\right)  <r<\eta\left(  \mu,\nu\right)  \cdot
R_{0}\left(  \nu\right)  .
\]
\end{definition}

\begin{remark}
For simplicity we could have asked the scaling property for $C_{0}\cdot
\eta\left(  \mu,\nu\right)  <r<r_{0}$ for a constant $r_{0}$ (a measure of the
integral scale). However, such a formulation could be too restrictive. On the
other hand, it is necessary that the range of $r$'s increases to infinity (relative to $\eta$) as
$\nu\rightarrow0$, otherwise the property becomes trivial, see remark
\ref{local property}.
\end{remark}

This is the mathematical formulation of K41 theory that we analyse in this
note. Here is a list of facts we can prove around it. In summary, they have
the structure of certain \textit{necessary conditions} for K41, and certain
almost \textit{equivalent conditions}.

\begin{itemize}
\item  We introduce a measure $\theta$ of the length scale where dissipation
takes place, defined as
\begin{equation}
\theta^{2}=\frac{\mu\left[  \int_{\left[  0,1\right]  ^{d}}\left\|  Du\left(
x\right)  \right\|  ^{2}dx\right]  }{\mu\left[  \int_{\left[  0,1\right]
^{d}}\left\|  D^{2}u\left(  x\right)  \right\|  ^{2}dx\right]  }\label{theta}%
\end{equation}
and prove a claim of the following form (theorem \ref{theorem on theta}):\ if
a scaling law holds on a range $C_{0}\cdot\eta<r<\eta\cdot R_{0}\left(
\nu\right)  $, then
\[
\theta\leq C\eta\;.
\]
In other words, the scale at which dissipation dominates cannot overlap with the range
over which a fractal scaling law holds.

\item  Since $\theta$ is constant (with respect to $\nu$)\ for both the 2D
stochastic Navier-Stokes equations and the Stokes (linear)\ equations, we can rule
out K41 scaling law for such systems. For the theory of 2D stochastic
Navier-Stokes equations this seems a remarkable fact. Moreover, these facts
tell us that, in case definition \ref{defk41} holds true in the 3D case, it is
strictly due to \textit{3D nonlinear} effects. This is further emphasised by
the following result.

\item  For the 3D stochastic Navier-Stokes equations we prove that, if K41
holds, then the \textit{mean vortex stretching}
\[
\mu\left[  \int_{\mathcal{T}}\left|\scal{S_{u}\Curl u,\Curl %
u}\right|  ^{2}dx\right]
\]
(where we set $S_{u}=\frac{1}{2}(  Du+Du^{T})  $) must be very large,
essentially at least as large as $\nu^{-3/2}$. See Corollaries~
\ref{stretching under additional hp} and~\ref{3Dsecondcorollary}. Vortex stretching is thus a
basic mechanism in K41 theory.

\item  We apply a known scaling transformation (see~\cite{Kup}) and introduce
an auxiliary family \ref{e:SNS} of stochastic Navier-Stokes equations with
modified domain and viscosity. Then we introduce a condition on this family of
equations, called \textit{Condition~\ref{condA}}, and prove it is equivalent to the
scaling law of K41 type. This is conceptually interesting since Condition~\ref{condA}
is of rather qualitative nature, while its consequence (the scaling law of K41
type) is more quantitative. More specifically, the behaviour $r^{2/3}$, and
also the exponent $3/4$ in the definition of $\eta$, arise from the stochastic
Navier-Stokes equations themselves, through a scaling transformation, when
certain bounds (without special exponents) are fulfilled for the auxiliary family.

\item  We give other necessary and sufficient conditions for K41,
starting from Condition~\ref{condA}. In plain words, they state that the large
structures of $u$ have bounded mean square gradients (or bounded mean enstrophy), while the small
structures have bounded mean energy.

\item  As a mild support to the belief that all these Conditions could be true
in dimension $d=3$, we finally exhibit a random field that satisfies them, and
was constructed independently from this purpose in~\cite{FlaGub} as a model of
turbulent fluid inspired by the vortex structures usually observed in
numerical simulations.
\end{itemize}

The remainder of the article is structured as follows. The remainder of the present section
introduces the notations that will be used throughout this work. In Section~\ref{sec:NecCond},
we draw several conclusions  from our formulation of the K41 scaling law that allow us to
give stringent necessary conditions for it to hold. These conditions are sufficient to rule out
any non-trivial scaling law in the 2D case. We proceed in Section~\ref{sec:NecSuf} to
find a condition that turns out to be equivalent to K41. This condition is then shown
in Section~\ref{sec:Eddy} to hold for a random eddy model introduced in \cite{FlaGub}.


\subsection{Notations about functions spaces\label{section 1.1}}

Let $\mathcal{T}$ be the torus $\left[  0,1\right]  ^{d}$, $d=2,3$,
$\mathbb{L}^{2}\left(  \mathcal{T}\right)  $ be the space of vector fields
$u:\mathcal{T}\rightarrow\mathbb{R}^{d}$ with $L^{2}\left(  \mathcal{T}%
\right)  $-components, $\mathbb{H}^{\alpha}\left(  \mathcal{T}\right)  $ be
the analogous Sobolev spaces, $\mathbb{C}\left(  \mathcal{T}\right)  $ be the
analogous space of continuous fields.

Let $H$ be the space of all fields $u\in\mathbb{L}^{2}\left(  \mathcal{T}%
\right)  $ such that $\Div u=0$ and $\int_{\mathcal{T}}u\left(  x\right)
dx=0$ (zero mean) and the trace of $u\cdot
n $ on the boundary is periodic (where $n$ is the outer normal, see~\cite{Temam}, Ch.I, Thm 1.2). 
Let $V$ be the
space of divergence free, zero mean, periodic elements of $\mathbb{H}%
^{1}\left(  \mathcal{T}\right)  $ and $D(A)$ be the space of divergence free,
zero mean, periodic elements of $\mathbb{H}^{2}\left(  \mathcal{T}\right)  $.
Finally, let $\mathcal{D}$ be the space of infinitely differentiable
divergence free, zero mean, periodic fields on $\mathcal{T}$. The spaces $V$,
$D(A)$ and $\mathcal{D}$ are dense and compactly embedded in $H$. 
Let $A:D(A)\subset H\rightarrow H$ be the (Stokes) operator
$Au=-\triangle u$ (componentwise).

Sometimes we shall also need the same framework for the torus $\left[
0,L\right]  ^{d}$, $d=2,3$, with any $L>0$. We set $\mathcal{T}_{L}=\left[
0,L\right]  ^{d}$, $H_{L}$ equal to the set of all fields $u\in\mathbb{L}%
^{2}\left(  \mathcal{T}_{L}\right)  $ such that $\Div u=0$ and $u\cdot
n $ on the boundary is periodic, $V_{L}$, $D\left(  A_{L}\right)  $ and
$A_{L}:D\left(  A_{L}\right)  \subset H_{L}\rightarrow H_{L}$ the analogs of
$V$, $D\left(  A\right)  $ and $A$. Notice only that we define the inner
product as
\[
\left|  u\right|  _{H_{L}}^{2}=\frac{1}{L^{d}}\int_{\mathcal{T}_{L}}\left|
u\left(  x\right)  \right|  ^{2}dx\;.
\]
(So that, roughly speaking, $\left|  u\right|  _{H_{L}}^{2}\sim\left|
u\left(  0\right)  \right|  ^{2}$ for homogeneous fields.)

\subsection{The class $\mathcal{P}$ of probability measures}

Let $\mathcal{P}_{0}$ be the family of all probability measures $\mu$ on $H$
(equipped with the Borel $\sigma$-algebra) such that $\mu\left(  D(A)\right)  =1$
($D(A)$ is a Borel set in $H$). Since $\mathbb{H}^{2}\left(  \mathcal{T}%
\right)  \subset\mathbb{C}\left(  \mathcal{T}\right)  $ by Sobolev embedding
theorem, the elements of $D(A)$ are continuous (have a continuous element in
their equivalence class). Consequently, given $x_{0}\in\mathcal{T}$, the
mapping $u\mapsto u\left(  x_{0}\right)  $ is well defined on $D(A)$, with
values in $\mathbb{R}^{d}$. In particular, any expression of the form
\[
\mu\left[  f\left(  u\left(  x_{1}\right)  ,...,u\left(  x_{n}\right)
\right)  \right]
\]
is well defined for given $x_{1},...,x_{n}\in\mathcal{T}$, given $\mu
\in\mathcal{P}_{0}$, and suitable $f:\mathbb{R}^{nd}\rightarrow\mathbb{R}$
(for instance measurable non negative). It follows that $S_{2}^{\mu}\left(
r\right)  $ is well defined (possibly infinite) for every $\mu\in
\mathcal{P}_{0}$.

The same argument does not apply to $Du\left(  x_{0}\right)  $ and
$D^{2}u\left(  x_{0}\right)  $, at least in $d=3$. This is why we use lengthy
expressions like
\[
\mu\left[  \int_{\mathcal{T}}\left\|  Du\left(  x\right)  \right\|
^{2}dx\right]  ,\quad\mu\left[  \int_{\mathcal{T}}\left\|  D^{2}u\left(
x\right)  \right\|  ^{2}dx\right]
\]
which are meaningful (possibly infinite) for every $\mu\in\mathcal{P}_{0}$.

We denote by $\mathcal{P}$ the class of all $\mu\in\mathcal{P}_{0}$ such that
\[
\mu\left[  \int_{\mathcal{T}}\left\|  Du\left(  x\right)  \right\|
^{2}dx\right]  <\infty
\]
and, for every $a\in\mathcal{T}$ and every rotation $R$ that transforms the set
of coordinate axes in itself,
\begin{equation}
\mu\left[  f\left(  u\left(  \cdot-a\right)  \right)  \right]  =\mu\left[
f\left(  u\right)  \right]  ,\quad\mu\left[  f\left(  u\left(  R\cdot\right)
\right)  \right]  =\mu\left[  f\left(  Ru(\cdot)\right)  \right]
\label{assumpintegr}%
\end{equation}
for all continuous bounded $f:H\rightarrow\mathbb{R}$. In plain words, we
impose space homogeneity and a discrete form of isotropy (compatible with the
symmetries of the torus). In the following we will refer to this symmetry as \emph{partial} or \emph{discrete} isotropy.

Discrete isotropy is imposed for two reasons. On one hand, $S_{2}^{\mu}\left(
r\right)  $ is independent of the coordinate unitary vector $e$, since given
two such vectors $e,e^{\prime}$ there is a rotation $R$ as above such that
$R\,e^{\prime}=e$, so
\begin{align*}
\mu\left[  \left\|  u\left(  re\right)  -u\left(  0\right)  \right\|
^{2}\right]   &=\mu\left[  \left\|  u\left(  R\,re^{\prime}\right)
-u\left(  R\,0\right)  \right\|  ^{2}\right] \\
& =\mu\left[  \left\|  R\left(  u\left(  re^{\prime}\right)  -u\left(
0\right)  \right)  \right\|  ^{2}\right] \\
&=  \mu\left[  \left\|  u\left(  re^{\prime}\right)  -u\left(  0\right)
\right\|  ^{2}\right]\;.
\end{align*}
On the other hand, we use discrete
isotropy through lemma \ref{lemmaisotropy} in Appendix~1. Finally, notice that $S_{2}^{\mu}(r)< \infty$ for every $r > 0$ and $\mu \in \mathcal{P}$, by Lemma~
\ref{lemmaTaylor} below.

\section{Necessary conditions for K41}
\label{sec:NecCond}

\subsection{General results}

The results of this subsection apply to suitable families of probability
measures, without any use of the Navier-Stokes equations. They will be applied
to the stochastic Navier-Stokes equations in the next subsection.

Given a measure $\mu\in\mathcal{P}$, $\mu\neq\delta_{0}$, we introduce the
number $\theta=\theta\left(  \mu\right)  $ defined by the identity
(\ref{theta}), letting $\theta=0$ when $\mu\left[
\int_{\mathcal{T}}\left\|  D^{2}u\left(  x\right)  \right\|
^{2}dx\right]  =\infty$. If $\mu=\delta_{0}$, numerator and denominator vanish
and we arbitrarily define $\theta=1$. We have $\theta\leq C$ where the
constant is universal and depends only on the Poincar\'{e}
constant of the torus. By the definitions, we have
\[
\theta\left(  \mu\right)  ^{2}=\frac{\epsilon\left(  \mu,\nu\right)  }%
{\nu\cdot\mu\left[  \int_{\mathcal{T}}\left\|  D^{2}u\left(  x\right)
\right\|  ^{2}dx\right]  }%
\]
for every pair $\left(  \mu,\nu\right)  \in\mathcal{P}\times\mathbb{R}_{+}$.

When, as in our application, the elements $u\in H$ have the meaning of
velocity fields, by dimensional analysis we see that $\theta$ has the
dimension of a length. We interpret it as \textit{an} \textit{estimate of the
length scale where dissipation is more relevant}. Indeed, very roughly, from
\[
\frac{\int_{\mathcal{T}}\left\|  D^{2}u\left(  x\right)  \right\|  ^{2}%
dx}{\int_{\mathcal{T}}\left\|  Du\left(  x\right)  \right\|  ^{2}dx}\sim
\frac{\sum\left|  k\right|  ^{2}\left(  \left|  k\right|  ^{2}\left|
\widehat{u}\left(  k\right)  \right|  ^{2}\right)  }{\sum\left|  k\right|
^{2}\left|  \widehat{u}\left(  k\right)  \right|  ^{2}}%
\]
we see that $\theta\left(  \mu\right)  ^{-2}$ has the meaning of typical
square wave length of dissipation (looking at $\left|  k\right|  ^{2}\left|
\widehat{u}\left(  k\right)  \right|  ^{2}$ as a sort of distribution in wave
space of the dissipation).

\begin{lemma}
\label{lemmaTaylor}For every $\mu\in\mathcal{P}$ such that $\theta\left(
\mu\right)  >0$ we have
\begin{equation}
\frac{1}{4d}\cdot r^{2}\leq\frac{S_{2}^{\mu}\left(  r\right)  }{\mu\left[
\int_{\mathcal{T}}\left\|  Du\left(  x\right)  \right\|  ^{2}dx\right]  }\leq
r^{2}\label{double inequality from Taylor}%
\end{equation}
for every $r\in(0,\theta(\mu)/{4d}]$. The upper bound is true for every $r > 0$ even if $\theta(\mu) = 0$.
\end{lemma}

\begin{proof}
Since we want to use Taylor formula for elements of $D(A)$, we use the
mollification described in Appendix 1. We denote by $\mu_{\varepsilon}$ the
mollifications of $\mu$. We prove in Appendix 1 that, for given $r$ and $\mu
$,
\[
\lim_{\varepsilon\rightarrow0}\mu_{\varepsilon}\left[  \left\|  Du\left(
0\right)  \right\|  ^{2}\right]  =\mu\left[  \int_{\mathcal{T}}\left\|
Du\left(  x\right)  \right\|  ^{2}dx\right]
\]%
\[
\lim_{\varepsilon\rightarrow0}\mu_{\varepsilon}\left[  \left\|  D^{2}u\left(
0\right)  \right\|  ^{2}\right]  =\mu\left[  \int_{\mathcal{T}}\left\|
D^{2}u\left(  x\right)  \right\|  ^{2}dx\right]
\]%
\[
\lim_{\varepsilon\rightarrow0}\mu_{\varepsilon}\left[  \left\|  u\left(
re\right)  -u\left(  0\right)  \right\|  ^{2}\right]  =\mu\left[  \left\|
u\left(  re\right)  -u\left(  0\right)  \right\|  ^{2}\right]  .
\]
By space homogeneity of $\mu_{\varepsilon}$
\begin{align*}
\mu_{\varepsilon}\left[  \left\|  u\left(  re\right)  -u\left(  0\right)
\right\|  ^{2}\right]   &  \leq r^{2}\int_{0}^{1}\mu_{\varepsilon}\left[
\left\|  Du\left(  \sigma e\right)  \right\|  ^{2}\right]  d\sigma\\
&  =r^{2}\mu_{\varepsilon}\left[  \left\|  Du\left(  0\right)  \right\|
^{2}\right]
\end{align*}
and thus, by the previous convergence results,
\[
\mu\left[  \left\|  u\left(  re\right)  -u\left(  0\right)  \right\|
^{2}\right]  \leq r^{2}\mu\left[  \int_{\mathcal{T}}\left\|  Du\left(
x\right)  \right\|  ^{2}dx\right]  .
\]
This implies the right-hand inequality of (\ref{double inequality from
Taylor}) for every $r>0$.

On the other hand, for smooth vector fields we have
\[
u\left(  re\right)  -u\left(  0\right)  =Du\left(  0\right)  re+r^{2}\int
_{0}^{1}D^{2}u\left(  \sigma e\right)  \left(  e,e\right)  d\sigma
\]
and thus
\begin{align*}
\mu_{\varepsilon}\left[  \left\|  Du\left(  0\right)  re\right\|
^{2}\right]   &  \leq2\mu_{\varepsilon}\left[  \left\|  u\left(  re\right)
-u\left(  0\right)  \right\|  ^{2}\right]  \\
&  +2\mu_{\varepsilon}\left[  \left\|  r^{2}\int_{0}^{1}D^{2}u\left(  \sigma
e\right)  \left(  e,e\right)  d\sigma\right\|  ^{2}\right]  .
\end{align*}
Again from space homogeneity of $\mu_{\varepsilon}$,
\[
\mu_{\varepsilon}\left[  \left\|  r^{2}\int_{0}^{1}D^{2}u\left(  \sigma
e\right)  \left(  e,e\right)  d\sigma\right\|  ^{2}\right]  \leq r^{4}%
\mu_{\varepsilon}\left[  \left\|  D^{2}u\left(  0\right)  \right\|
^{2}\right]
\]
and from lemma \ref{lemmaisotropy} of Appendix 1
\[
\mu_{\varepsilon}\left[  \left\|  Du\left(  0\right)  e\right\|  ^{2}\right]
=\frac{1}{d}\mu_{\varepsilon}\left[  \left\|  Du\left(  0\right)  \right\|
^{2}\right]  .
\]
Therefore
\[
\mu_{\varepsilon}\left[  \left\|  u\left(  re\right)  -u\left(  0\right)
\right\|  ^{2}\right]  \geq\frac{r^{2}}{2d}\mu_{\varepsilon}\left[  \left\|
Du\left(  0\right)  \right\|  ^{2}\right]  -r^{4}\mu_{\varepsilon}\left[
\left\|  D^{2}u\left(  0\right)  \right\|  ^{2}\right]  .
\]
We thus have in the limit
\[
S_{2}\left(  r\right)  \geq\frac{r^{2}}{2d}\mu\left[  \int_{\mathcal{T}%
}\left\|  Du\left(  x\right)  \right\|  ^{2}dx\right]  -r^{4}\mu\left[
\int_{\mathcal{T}}\left\|  D^{2}u\left(  x\right)  \right\|  ^{2}dx\right]
\]
and therefore, by definition of $\theta\left(  \mu\right)  $,
\[
S_{2}\left(  r\right)  \geq\left(  \frac{1}{2d}-\frac{r^{2}}{\theta\left(
\mu\right)  }\right)  \mu\left[  \int_{\mathcal{T}}\left\|  Du\left(
x\right)  \right\|  ^{2}dx\right]  \cdot r^{2}.
\]
This implies the left-hand inequality of (\ref{double inequality from Taylor})
for $r\in(0,\frac{\theta\left(  \mu\right)  }{4d}]$. The proof is complete.
\end{proof}

\begin{theorem}
\label{theorem on theta}Let $\mathcal{M}\subset\mathcal{P}\times\mathbb{R}%
_{+}$ be a set with the following scaling property:\ there is a function
$\widetilde{\eta}:\mathcal{M}\rightarrow\mathbb{R}_{+}$ (the length scale of
the scaling property), a decreasing function $R_{0}:[0,\infty)\rightarrow
\mathbb{R}_{+}$, with $\lim_{\nu\rightarrow0}R_{0}\left(  \nu\right)
=+\infty$, a scaling exponent $\alpha\in\left(  0,2\right)  $ and constants
$C_{2}\geq C_{1}>0$, $C_{3}>0$, $\nu_{0}>0$, such that $R_{0}\left(
\nu\right)  >C_{3}$ and
\begin{equation}
C_{1}\cdot r^{\alpha}\leq S_{2}^{\mu}\left(  r\right)  \leq C_{2}\cdot
r^{\alpha}\text{ \quad for }r\in\left[  C_{3}\widetilde{\eta}\left(  \mu
,\nu\right)  ,\widetilde{\eta}\left(  \mu,\nu\right)  R_{0}\left(  \nu\right)
\right]  \label{scaling}%
\end{equation}
for every $\nu\in\left(  0,\nu_{0}\right)  $ and every $\mu\in\mathcal{M}%
_{\nu}$. Let $\theta\left(  \mu\right)  $ be the dissipation length scale
defined above.

Then the two length scales $\theta\left(  \mu\right)  $ and $\widetilde{\eta
}\left(  \mu,\nu\right)  $ are related by the property
\begin{equation}
\underset{\nu\rightarrow0}{\lim\sup}\left(  \sup_{\mu\in\mathcal{M}_{\nu}%
}\frac{\theta\left(  \mu\right)  }{\widetilde{\eta}\left(  \mu,\nu\right)
}\right)  <\infty.\label{comparison}%
\end{equation}
\end{theorem}

\begin{proof}
It is intuitively rather clear that (\ref{double inequality from Taylor}) is
in contradiction with (\ref{scaling}) if the ranges of $r$ where the two
properties hold overlap, so we need the bound (\ref{comparison}). The proof below confirm this intuition by ruling out the possibility that the factor $\mu\left[
\int_{\mathcal{T}}\left\|  Du\left(  x\right)  \right\|  ^{2}dx\right]  $ may
produce a compensation.

We argue by contradiction and assume that there exists a sequence $\left(
\mu_{n},\nu_{n}\right)  \in\mathcal{M}$, with $\nu_{n}\rightarrow0$, such
that
\begin{equation}
\lim_{n\rightarrow\infty}\frac{\theta\left(  \mu_{n}\right)  }{\eta\left(
\mu_{n},\nu_{n}\right)  }=+\infty.\label{contradiction}%
\end{equation}
Notice that, in such a case, $\theta\left(  \mu_{n}\right)  $ must be
positive, so lemma \ref{lemmaTaylor} applies. Let us consider two sequences
$r_{n}^{\prime}$ and $r_{n}^{\prime\prime}$ defined as follows:
\[
r_{n}^{\prime}=C_{3}\eta\left(  \mu_{n},\nu_{n}\right)  ,\quad r_{n}%
^{\prime\prime}=r_{n}^{\prime}a_{n}%
\]
with
\[
\lim_{n\rightarrow\infty}a_{n}=+\infty,\quad r_{n}^{\prime\prime}\leq
\eta\left(  \mu_{n},\nu_{n}\right)  R_{0}\left(  \nu_{n}\right)  ,\quad
r_{n}^{\prime\prime}\leq\frac{\theta\left(  \mu_{n}\right)  }{4d}%
\]
where we ask that the last two inequalities are satisfied at least eventually.
Such a sequence $r_{n}^{\prime\prime}$ exists because $\lim_{\nu\rightarrow
0}R_{0}\left(  v\right)  =+\infty$ and (\ref{contradiction}) is assumed.

We have (eventually) $r_{n}^{\prime}$, $r_{n}^{\prime\prime}\in(0,\frac
{\theta\left(  \mu_{n}\right)  }{4d}]$ and $r_{n}^{\prime}$, $r_{n}%
^{\prime\prime}\in\left[  C_{3}\eta\left(  \mu_{n},\nu_{n}\right)
,\eta\left(  \mu_{n},\nu_{n}\right)  R_{0}\left(  \nu_{n}\right)  \right]  $,
hence for both $r_{n}:=r_{n}^{\prime}$ and $r_{n}:=r_{n}^{\prime\prime}$ we
have
\[
C_{1}r_{n}^{\alpha}\leq S_{2}^{\mu_{n}}\left(  r_{n}\right)  \leq C_{2}%
r_{n}^{\alpha},\quad\frac{1}{4d}\beta_{n}r_{n}^{2}\leq S_{2}^{\mu_{n}}\left(
r_{n}\right)  \leq\beta_{n}r_{n}^{2}%
\]
where we have set $\beta_{n}=\mu_{n}\left[  \int_{\mathcal{T}}\left\|
Du\left(  x\right)  \right\|  ^{2}dx\right]  $. The contradiction will come
from the fact that, if it could happen that $\beta_{n}$ adjusts the factor
$r_{n}^{2}$ to produce $r_{n}^{\alpha}$, this cannot happen simultaneously for
the two sequences $r_{n}=r_{n}^{\prime}$ and $r_{n}=r_{n}^{\prime\prime}$.
Indeed, from the previous inequalities we must have
\[
C_{1}r_{n}^{\alpha}\leq\beta_{n}r_{n}^{2},\quad\beta_{n}r_{n}^{2}\leq
4dC_{2}r_{n}^{\alpha}%
\]
hence
\[
\beta_{n}\geq C_{1}r_{n}^{\alpha-2},\quad\beta_{n}\leq4dC_{2}r_{n}^{\alpha-2}%
\]
for both $r_{n}=r_{n}^{\prime}$ and $r_{n}=r_{n}^{\prime\prime}$. But the
inequalities
\[
\beta_{n}\geq C_{1}\left(  r_{n}^{\prime}\right)  ^{\alpha-2},\quad\beta
_{n}\leq4dC_{2}\left(  r_{n}^{\prime\prime}\right)  ^{\alpha-2}%
\]
and the assumption $\alpha<2$ imply
\[
r_{n}^{\prime}\geq Cr_{n}^{\prime\prime}%
\]
eventually, for a suitable constant $C>0$. This is impossible since
$\lim_{n\rightarrow\infty}a_{n}=+\infty$. The proof is complete.
\end{proof}

\begin{remark}
\label{local property}The divergent factor $R_{0}\left(  \nu\right)  $ in the
definition (\ref{scaling}) of a scaling law is essential to have a non trivial
definition. If, on the contrary, we simply ask that the scaling law holds on a
bounded interval $r\in\left[  C_{3}\eta_{\nu},C_{4}\eta_{\nu}\right]  $, we
have a definition without real interest. Let us explain this fact with a
(useless) definition and an example. Let us say that a family $\mathcal{M}%
\subset\mathcal{P}\times\mathbb{R}_{+}$ satisfies a local $\alpha$ property,
$\alpha<2$, if\ there is a function $\widetilde{\eta}\left(  \mu,\nu\right)  $
and constants $C_{2}\geq C_{1}>0$, $C_{4}\geq C_{3}>0$, $\nu_{0}>0$, such
that
\begin{equation}
C_{1}r^{\alpha}\leq S_{2}^{\mu}\left(  r\right)  \leq C_{2}r^{\alpha}\text{
\quad for }r\in\left[  C_{3}\widetilde{\eta}\left(  \mu,\nu\right)
,C_{4}\widetilde{\eta}\left(  \mu,\nu\right)  \right]  \label{localscaling}%
\end{equation}
for every $\nu\in\left(  0,\nu_{0}\right)  $ and every $\mu\in\mathcal{M}%
_{\nu}$. As an example, consider a case with the mapping $\nu\longmapsto$
$\mathcal{M}_{\nu}$ which is single valued and injective and
\[
S_{2}^{\mu^{\nu}}\left(  r\right)  =\nu^{-1}r^{2}%
\]
where $\mathcal{M}_{\nu}=\left\{  \mu^{\nu}\right\}  $. This function
$S_{2}^{\mu^{\nu}}\left(  r\right)  $ certainly does not have any interesting
scaling exponent (different from 2) but satisfies the previous local $\alpha$
property simultaneously for a continuum of values of $\alpha$. Indeed, given
any $\alpha\in\left(  0,2\right)  $ take $\widetilde{\eta}\left(  \mu^{\nu
},\nu\right)  =\nu^{\frac{1}{2-\alpha}}$; then given a choice of $C_{4}\geq
C_{3}>0$, for every $r\in\left[  C_{3}\widetilde{\eta}\left(  \mu^{\nu}%
,\nu\right)  ,C_{4}\widetilde{\eta}\left(  \mu^{\nu},\nu\right)  \right]  $,
namely for $\nu^{-\frac{1}{2-\alpha}}r\in\left[  C_{3},C_{4}\right]  $, we
have
\[
S_{2}^{\mu^{\nu}}\left(  r\right)  =\left(  \nu^{-\frac{1}{2-\alpha}}r\right)
^{2-\alpha}r^{\alpha}\in\left[  C_{1},C_{2}\right]  \cdot r^{\alpha}%
\]
with $C_{1}=C_{3}^{2-\alpha}$, $C_{2}=C_{4}^{2-\alpha}$. This example shows
that the local $\alpha$ property is not a distinguished scaling property.
Moreover, it shows that (\ref{double inequality from Taylor}) and
(\ref{localscaling}) are compatible: this is why a proof of theorem
\ref{theorem on theta} is necessary.
\end{remark}

\begin{example}
Since we have just given a negative example (artificial, but close to what
happens in 2D), let us also give an example of a function of $\left(
\nu,r\right)  $ which satisfies the properties of definition \ref{defk41} and
also \ref{double inequality from Taylor} (to see that they are compatible). It
may look artificial, but it was devised on the basis of the vortex model of
\cite{FlaGub}, described also below. The function is
\[
S_{2}^{\mu^{\nu}}\left(  r\right)  =\int_{\eta}^{1}l^{2/3}\left(
\frac{l\wedge r}{l}\right)  ^{2}\frac{dl}{l}%
\]
with $\eta=\nu^{3/4}$. We have
\[
r\leq\eta\Rightarrow S_{2}^{\mu^{\nu}}\left(  r\right)  =\int_{\eta}%
^{1}l^{2/3}\left(  \frac{r}{l}\right)  ^{2}\frac{dl}{l}=\frac{3}{4}%
r^{2}\left[  \nu^{-1}-1\right]
\]
which is essentially the behaviour \ref{double inequality from Taylor}. On the
other hand,
\begin{align*}
r  & \in\left[  \eta,1\right]  \Rightarrow S_{2}^{\mu^{\nu}}\left(  r\right)
=\int_{\eta}^{r}l^{2/3}\frac{dl}{l}+\int_{r}^{1}l^{2/3}\left(  \frac{r}%
{l}\right)  ^{2}\frac{dl}{l}\\
& =\frac{9}{4}r^{2/3}-\frac{3}{2}\nu^{1/2}-\frac{3}{4}r^{2}%
\end{align*}
which is bounded above and below by the order $r^{2/3}$ since $r\in\left[
\nu^{3/4},1\right]  $ ($\nu^{1/2}\leq r^{2/3}$).
\end{example}

Let us finally state two general consequences of the previous theorem, that we
shall apply to stochastic Navier-Stokes equations.

\begin{corollary}
Given a family $\mathcal{M}\subset\mathcal{P}\times\mathbb{R}_{+}$, \bigskip
if
\[
\inf_{\left(  \mu,\nu\right)  \in\mathcal{M}}\theta\left(  \mu\right)
>0\text{ }%
\]
then no scaling law in the sense of the previous theorem may hold true with a
length scale $\widetilde{\eta}\left(  \mu,\nu\right)  $ such that
\[
\underset{\nu\rightarrow0}{\lim\inf}\left(  \inf_{\mu\in\mathcal{M}_{\nu}%
}\widetilde{\eta}\left(  \mu,\nu\right)  \right)  =0.
\]
\end{corollary}

We shall see that this simple corollary applies to the 2D stochastic
Navier-Stokes equation and the Stokes problem, so K41 scaling law is ruled out
for these systems.

Let us apply the theorem to the case of K41 scaling law. We take, in the
previous theorem,
\[
\widetilde{\eta}\left(  \mu,\nu\right)  =\eta\left(  \mu,\nu\right)
=\nu^{3/4}\epsilon\left(  \mu,\nu\right)  ^{-1/4}%
\]
as in the introduction. In the following result, $\mu\left[  \int
_{\mathcal{T}}\left\|  D^{2}u\left(  x\right)  \right\|  ^{2}dx\right]  $ may
be infinite.

\begin{corollary}
Let $\mathcal{M}\subset\mathcal{P}\times\mathbb{R}_{+}$ be a family with the
K41 scaling law, in the sense of Definition~\ref{defk41}. Then there exist
$\nu_{0}>0$ and $C>0$ such that
\[
\mu\left[  \int_{\mathcal{T}}\left\|  D^{2}u\left(  x\right)  \right\|
^{2}dx\right]  \geq C\epsilon^{3/2}\left(  \mu,\nu\right)  \cdot\nu^{-5/2}%
\]
for every $\nu\in\left(  0,\nu_{0}\right)  $ and every $\mu\in\mathcal{M}%
_{\nu}$.
\end{corollary}

\begin{proof}
From (\ref{comparison}), the definition of $\eta\left(  \mu,\nu\right)  $ and
the definition of $\theta^{2}\left(  \mu\right)  $ we have
\[
\underset{\nu\rightarrow0}{\lim\sup}\left(  \sup_{\mu\in\mathcal{M}_{\nu}%
}\frac{\mu\left[  \int_{\mathcal{T}}\left\|  Du\left(  x\right)  \right\|
^{2}dx\right]  }{\nu^{3/2}\epsilon\left(  \mu,\nu\right)  ^{-1/2}\mu\left[
\int_{\mathcal{T}}\left\|  D^{2}u\left(  x\right)  \right\|  ^{2}dx\right]
}\right)  <\infty.
\]
Thus, from the definition of $\epsilon\left(  \mu,\nu\right)  $,
\[
\underset{\nu\rightarrow0}{\lim\sup}\left(  \sup_{\mu\in\mathcal{M}_{\nu}%
}\frac{\nu^{-5/2}\epsilon\left(  \mu,\nu\right)  ^{3/2}}{\mu\left[
\int_{\mathcal{T}}\left\|  D^{2}u\left(  x\right)  \right\|  ^{2}dx\right]
}\right)  <\infty.
\]
This implies the claim of the Corollary.
\end{proof}

\begin{remark}
Dimensional analysis says that $\nu$ has dimension $\left[  L\right]
^{2}\left[  T\right]  ^{-1}$, $\epsilon$ has dimension $\left[  L\right]
^{2}\left[  T\right]  ^{-3}$, so $\epsilon^{3/2}\left(  \mu,\nu\right)
\cdot\nu^{-5/2}$ has dimension $\left[  L\right]  ^{-2}\left[  T\right]
^{-2}$, the correct dimension of $E^{\mu}\left[  \int_{\mathcal{T}}\left\|
D^{2}u\left(  x\right)  \right\|  ^{2}dx\right]  $.
\end{remark}


\subsection{Application to stochastic Navier-Stokes equations}

In this section we consider equation (\ref{stochNS}) in dimension 2 and 3 and
also the corresponding linear equations (Stokes equations).

\subsubsection{The noise\label{subsubsectnoise}}

Since we are dealing with spaces of translation invariant measures, we wish
to consider classes of noises that produce such measures. Every Gaussian
translation invariant noise is `diagonal' with respect to the Stokes operator $A$
in the sense that eigenmodes are all independent. In order to give a rigorous
definition for our driving noise, we define
\[
\Lambda^{(\infty)}:=\left\{k\in 2\pi\mathbb{Z}^d\ :\ |k|>0\right\}
\]
and we assume that the noise of equation (\ref{stochNS}) has the form
\begin{equation}\label{formnoise}
\sum_{k\in\Lambda^{(\infty)}}\sigma_k{\dot\beta}_k(t)
e^{-ik\cdot x}
\end{equation}
where $(\beta_k)_{k\in\Lambda^{(\infty)}}$ are independent $d$-dimensional
Brownian motions and $(\sigma_k)_{k\in\Lambda^{(\infty)}}$ are $d\times d$
complex-valued matrices such that
\begin{equation}\label{ip1noise}
k\cdot\sigma_k=0
\end{equation}
and
\begin{equation}\label{ip2noise}
\sum_{k\in\Lambda^{(\infty)}}|\sigma_k|^2<\infty.
\end{equation}
Moreover, in order to obtain real-valued noise, we assume that
\begin{equation}
\overline{\sigma_k}=\sigma_{-k}\label{ip3noise}
\end{equation}
for every $k\in\Lambda^{(\infty)}$. Additionally, the vector-valued random field
\[
W(t,x)=\sum_{k\in\Lambda^{(\infty)}}\sigma_k\beta_k(t)e^{-ik\cdot x}
\]
is, for every $t\ge0$, partially isotropic if and only 
\begin{equation}\label{ip4noise}
|\sigma_k|=|\sigma_{Rk}|
\end{equation}
for all $k\in\Lambda^{(\infty)}$ and
for every coordinate rotation $R$.

Finally, in order to have measures with $\mu\left(  D(A)\right)  =1$ we assume that
\begin{equation}
\sum_{k\in\Lambda^{(\infty)}}|k|^2|\sigma_k|^2<\infty,
\label{ip5noise}%
\end{equation}
since the values $|k|^2$ correspond to the eigenvalues of $A$.
To summarise, we shall always assume that the
noise (\ref{formnoise}) satisfies assumptions (\ref{ip1noise})-(\ref{ip5noise}).

\subsubsection{The two-dimensional case}

The following result is well known.

\begin{lemma}
Let $\mu$ be an invariant measure of (\ref{stochNS}) ($d=2$) such that
\[
\mu\int_{\mathcal{T}}\left\|  Du\left(  x\right)  \right\|  ^{2}dx<\infty.
\]
Then $\mu\in\mathcal{P}_{0}$ and
\[
\nu\cdot\mu\int_{\mathcal{T}}\left\|  Du\left(  x\right)  \right\|
^{2}dx=\frac{1}{2}\sum_{k\in\Lambda^{\left(
\infty\right)  }}\left|  \sigma_k\right|  ^{2}%
\]%
\[
\nu\cdot\mu\int_{\mathcal{T}}\left\|  D\Curl u\left(  x\right)
\right\|  ^{2}dx=\frac{1}{2}\sum_{k\in\Lambda^{\left(
\infty\right)  }}\left|  k\right|  ^{2}\left|  \sigma_k\right|  ^{2}.
\]
\end{lemma}

\begin{proof}

Given $\mu$, consider the (product) filtered probability space $(
\Omega,\mathcal{A},(  \mathcal{A}_{t})  _{t\geq0},P)  $
supporting both a family of independent $d$-dimensional Brownian motions $\beta_{k}\left(  t\right)  $, $\left(  k,\alpha\right)  \in\Lambda^{\left(
\infty\right)  }$, and a non anticipating random variable $u_{0} \in \mathcal{A}_{0}$ with law
$\mu$. The corresponding strong solution $u(t,x)$ of (\ref{stochNS}) is a
stationary process and satisfies, due to It\^{o} formula, the balance
relations
\begin{align*}
& \frac{1}{2}E^{P}\int_{\mathcal{T}}\left\|  u\left(  t,x\right)  \right\|
^{2}dx+\nu E^{P}\int_{0}^{t}\int_{\mathcal{T}}\left\|  Du\left(  s,x\right)
\right\|  ^{2}dx\\
& =\frac{1}{2}E^{P}\int_{\mathcal{T}}\left\|  u_{0}\left(  x\right)  \right\|
^{2}dx+\frac{1}{2}\sum_{k\in\Lambda^{(
\infty)}}\left|  \sigma_k\right|  ^{2}\cdot t
\end{align*}%
\begin{align*}
& \frac{1}{2}E^{P}\int_{\mathcal{T}}\left\|  \Curl u\left(  t,x\right)
\right\|  ^{2}dx+\nu E^{P}\int_{0}^{t}\int_{\mathcal{T}}\left\|
D\Curl u\left(  s,x\right)  \right\|  ^{2}dx\\
& =\frac{1}{2}E^{P}\int_{\mathcal{T}}\left\|  \Curl u_{0}\left(
x\right)  \right\|  ^{2}dx+\frac{1}{2}\sum_{k\in\Lambda^{(\infty)}}
\left|  k\right|  ^{2}\left|
\sigma_k\right|  ^{2}\cdot t.
\end{align*}
The result easily follows from stationarity.
\end{proof}

\begin{corollary}
There exists a positive constant $\theta_{0}$, independent of $\nu$, such
that
\[
\theta\left(  \mu\right)  \geq\theta_{0}%
\]
for every invariant measure $\mu\in\mathcal{P}$ of (\ref{stochNS}).
\end{corollary}

\begin{proof}
The property $\theta\left(  \mu\right)  \geq\theta_{0}$ follows from the
definition of $\theta\left(  \mu\right)  $ and the two identities of the
previous lemma, since
\[
\int_{\mathcal{T}}\left\|  D^{2}u\left(  x\right)  \right\|  ^{2}dx\leq
C\int_{\mathcal{T}}\left\|  D\Curl u\left(  x\right)  \right\|  ^{2}dx
\]
for a universal constant $C>0$.
\end{proof}

In the next theorem, when we say that $\mathcal{M}\subset\mathcal{P}%
\times\mathbb{R}_{+}$ is a family of invariant measures of (\ref{stochNS}), we
clearly understand that each element $\left(  \mu,\nu\right)  \in\mathcal{M} $
has the property that $\mu$ is an invariant measure for the Markov semigroup
associated to equation (\ref{stochNS}) with viscosity equal to $\nu$.

\begin{theorem}
In dimension $d=2$, a family of invariant measures $\mathcal{M}\subset
\mathcal{P}\times\mathbb{R}_{+}$ of (\ref{stochNS}) cannot have any scaling
law (in the sense of (\ref{scaling})).
\end{theorem}

\begin{remark}
Under our assumptions on the noise, invariant measures of (\ref{stochNS}) that
belong to $\mathcal{P}$ certainly exist. In principle there could exist invariant measures
for (\ref{stochNS}) not belonging to $\mathcal{P}$, but this has recently been excluded under very weak
conditions on the driving noise (see~\cite{HairMatt} and the
references therein).
\end{remark}

\begin{remark}
Consider equation (\ref{stochNS}) without the nonlinear term (called Stokes
equations):
\[
\frac{\partial u}{\partial t}+\nabla p=\nu\triangle u+\sum_{
k\in\Lambda^{(\infty)}}\sigma_k {\dot\beta}_k(t)
e^{-ik\cdot x}
\]
in dimension $d=2,3$. Let $\mathcal{M}\subset\mathcal{P}\times\mathbb{R}_{+}$
be a family of invariant measures for it. Then the same results of the
previous theorem hold true. The proof is the same. Alternatively, one may work
componentwise in the Fourier modes and prove easily the claims.
\end{remark}

\subsubsection{The three-dimensional case}

\label{subsub3D}

The lack of knowledge about the well posedness of the 3D stochastic
Navier-Stokes equations has, among its consequences, the absence of the Markov
property, and therefore of the usual notion of invariant measure. One may
introduce several variants. Here we adopt the following concept.

Consider the usual Galerkin approximations, recalled in Appendix B. The
equation with generic index $n$ in this scheme defines a Markov process, with
the Feller property, and has invariant measures, by the classical
Krylov-Bogoliubov method: if $X_{n}^{x}\left(  t\right)  $ is its solution
starting from $x$ and $\nu_{t}^{n,x}$ is the law of $X_{n}^{x}\left(
t\right)  $ on $H$, by It\^{o} formula it is easy to get a bound of the form
(see for instance~\cite{FlaGat})
\[
\sup_{T\geq0}\frac{1}{T}\int_{0}^{T}E\left[  \left\|  X_{n}^{x}\left(
t\right)  \right\|  _{V}^{2}\right]  dt\leq C<\infty
\]
which implies (\cite{ChowH} have been the first ones to use this elegant fast
method) the necessary tightness in $T$ of the time averaged measures
\[
\mu_{T}^{n,x}:=\frac{1}{T}\int_{0}^{T}\nu_{t}^{n,x}dt.
\]
If we choose the initial condition $x=0$, then $\mu_{T}^{n,x}\in\mathcal{P} $
(in particular it is space homogeneous and partially isotropic), so there exist
invariant measures in $\mathcal{P}$ for the Galerkin equation. Denote by
$\mathcal{S}^{n}$ the set of all such invariant measures (thus $\mathcal{S}%
^{n}\subset\mathcal{P}$).

The constant $C$ in the estimate above is also independent of $n$;\ it follows
that the invariant measures of the class $\mathcal{S}^{n}$ just constructed
fulfill the bound
\[
\mu^{n}\left[  \left\|  \cdot\right\|  _{V}^{2}\right]  \leq C.
\]
In fact it is possible to show that every element of $\mathcal{S}^{n}$ has
this property, \cite{FlaCetraro} (if we do not want to use this property, it
is sufficient to restrict the definition of $\mathcal{S}^{n}$ in the sequel).
These facts imply that $\cup_{n}\mathcal{S}^{n}$ is relatively compact in the
weak topology of probability measures on $H$. We denote by $\mathcal{P}%
_{{\mathrm{NS}}}^{G}\left(  \nu\right)  $ (the superscript $G$ will remind us
that we use the particular procedure of Galerkin approximations) the set of
limit points of $\cup_{n}\mathcal{S}^{n}$, precisely defined as follows: a
probability measure $\mu$ on $H$ belongs to $\mathcal{P}_{{\mathrm{NS}}}%
^{G}\left(  \nu\right)  $ if there is a sequence $k_{n}\rightarrow\infty$ and
elements $\mu_{k_{n}}\in\mathcal{S}^{k_{n}}$ such that $\mu_{k_{n}}$ converges
to $\mu$ in the weak topology of probability measures on $H$. The elements of
the set $\mathcal{P}_{{\mathrm{NS}}}^{G}\left(  \nu\right)  $ are space
homogeneous and partially isotropic (these relations are stable under weak
convergence). Furthermore, they have the other regularity properties required
to belong to $\mathcal{P}$:\ finite second moment in $V$ comes from the
previous estimates, $\mu\left(  D(A)\right)  =1$ from a regularity result
of~\cite{DaPDeb}, see also \cite{FlaCetraro}, summarized in the following
lemma. Therefore $\mathcal{P}_{{\mathrm{NS}}}^{G}\left(  \nu\right)
\subset\mathcal{P}$.

\begin{lemma}
Given $\nu>0$, there is a constant $C_{\nu}>0$ (depending on $\nu$) such that
\[
\mu_{n}\left(  \left|  A\cdot\right|  _{H}^{2/3}\right)  \leq C
\]
for every $n$ and every invariant measure $\mu_{n}\in\mathcal{S}^{n}$.
\end{lemma}

Given $u\in V$, let $S_{u}$ be the tensor with $L^{2}\left(  \mathcal{T}%
\right)  $ components
\[
S_{u}=\frac{1}{2}\left(  Du+Du^{T}\right)
\]
(called stress tensor). The scalar field
\[
\scal{S_{u}\left( x\right) \mathsf{curl}u\left( x\right) , \mathsf{curl}%
u\left( x\right)}%
\]
describes the stretching of the vorticity field. If we set $\xi=\mathsf{curl}%
u$, then formally we have
\[
\frac{\partial\xi}{\partial t}+\left(  u\cdot\nabla\right)  \xi=\nu
\triangle\xi+S_{u}\xi+i\sum_{k\in\Lambda^{(\infty)}}k\times\sigma_{k}%
\overset{\cdot}{\beta}_{k}e^{-ik\cdot x}.
\]
A \textit{formal} application of It\^{o} formula yields the inequality
\begin{align*}
\nu\cdot\mu\int_{\mathcal{T}}\left\|  D\mathsf{curl}u\left(  x\right)
\right\|  ^{2}dx &  \leq\mu\int_{\mathcal{T}}\scal{S_{u}\left( x\right
) \mathsf{curl}u\left( x\right), \mathsf{curl}u\left( x\right)}\,dx\\
&  +\frac{1}{2}\sum_{k\in\Lambda^{(\infty)}}\left|  k\right|  ^{2}\left|
\sigma_{k}\right|  ^{2}.
\end{align*}
for $\mu\in\mathcal{P}_{{\mathrm{NS}}}^{G}\left(  \nu\right)  $ (in fact
formally the identity). Along with the general results of the previous
sections we would get
\begin{equation}
\mu\left[  \int_{\mathcal{T}}\scal{S_{u}\left( x\right) \mathsf{curl}%
u\left( x\right), \mathsf{curl}u\left( x\right)}dx\right]  \geq C\epsilon
^{3/2}\left(  \mu,\nu\right)  \cdot\nu^{-3/2}.\label{desiderata}%
\end{equation}
This would be the final result of this section, having an interesting physical
interpretation. However we are not able to prove it in this form. We analyze the status of this inequality by presenting some related rigorous results. They are
of two different natures: Corollary~\ref{stretching for Galerkin} reformulates
it for the coarse graining scheme given by Galerkin approximations;\ Corollary~\ref{stretching under additional hp} expresses the most natural statement
directly for $\mu\in\mathcal{P}_{{\mathrm{NS}}}^{G}\left(  \nu\right)  $ but
it requires an additional unproved regularity assumption.

\begin{lemma}
Given $\mu\in\mathcal{P}_{{\mathrm{NS}}}^{G}\left(  \nu\right)  $, and
$\mu_{n_{k}}\in\mathcal{S}^{k_{n}}$ such that $\mu_{k_{n}}$ converges to $\mu$
in the weak topology of probability measures on $H$, then
\[
\mu\left[  \left|  A\cdot\right|  _{H}^{2}\right]  \leq\underline{\lim}%
\mu_{n_{k}}\left[  \left|  A\cdot\right|  _{H}^{2}\right]  .
\]
The same is true for $\mu\int_{\mathcal{T}}\left\|  D\mathsf{curl}u\left(
x\right)  \right\|  ^{2}dx$ in place of $\mu\left[  \left|  A\cdot\right|
_{H}^{2}\right]  $.
\end{lemma}

\begin{proof}
Let $\left\{  \varphi_{m}\right\}  _{m\in\mathbb{N}}\in C_{b}\left(  H\right)
$ be a sequence that converges monotonically increasing to $\left|
A\cdot\right|  _{H}^{2}$ for every $x\in D(A)$, it is easy to construct it by
cut-off and finite dimensional approximations). Since $\mu\left(  D(A)\right)
=1$, by Beppo-Levi theorem $\mu[  \varphi_{m}]  \rightarrow
\mu [  |  A\cdot |  _{H}^{2}]  $. Given $\varepsilon>0$,
let $m_{0}$ be such that $\mu[  \varphi_{m_{0}}]  \geq\mu[
|  A\cdot |  _{H}^{2} ]  -\varepsilon$. Since $\mu_{n_{k}%
}[  \varphi_{m_{0}}]  \rightarrow \mu[
\varphi_{m_{0}}]  $ as $k \to \infty$, eventually in $k$ we thus have $\mu_{n_{k}}[
\varphi_{m_{0}}]  \geq\mu[  |  A\cdot |  _{H}^{2} ]
-2\varepsilon$, and therefore also $\mu_{n_{k}}[  |  A\cdot |
_{H}^{2}]  \geq\mu[  |  A\cdot |  _{H}^{2}]
-2\varepsilon$. This proves the first part of the lemma; the second one is similar.
\end{proof}

\begin{corollary}
\label{stretching for Galerkin}Let $\mathcal{M}\subset\mathcal{P}%
\times\mathbb{R}_{+}$, with $\mathcal{M}_{\nu}\subset\mathcal{P}%
_{{\mathrm{NS}}}^{G}\left(  \nu\right)  $, be a family with the K41 scaling
law, in the sense of definition \ref{defk41}. Then there exist $\nu_{0}>0$
and $C>0$ such that
\[
\underset{k\rightarrow\infty}{\lim\inf}\mu_{n_{k}}\left[  \int_{\mathcal{T}%
}\scal{S_{u}\left( x\right) \mathsf{curl}u\left( x\right), \mathsf{curl}%
u\left( x\right)}dx\right]  \geq C\epsilon^{3/2}\left(  \mu,\nu\right)
\cdot\nu^{-3/2}%
\]
for every $\nu\in\left(  0,\nu_{0}\right)  $, every $\mu\in\mathcal{M}_{\nu}$
and every sequence $\mu_{n_{k}}\in\mathcal{S}^{k_{n}}$ such that $\mu_{k_{n}}$
converges to $\mu$ in the weak topology of probability measures on $H$.
\end{corollary}

\begin{proof}
From the previous section we know that
\[
\mu\int_{\mathcal{T}}\left\|  D^{2}u\left(  x\right)  \right\|  ^{2}%
dx\geq\epsilon^{3/2}\left(  \mu,\nu\right)  \cdot\nu^{-5/2}.
\]
Since
\begin{equation}
\left\langle Af,g\right\rangle _{H}=\left\langle \mathsf{curl}f,\mathsf{curl}%
g\right\rangle _{H}\label{A - curl}%
\end{equation}
for every $f,g\in D(A)$, we have
\[
\mu\int_{\mathcal{T}}\left\|  D\mathsf{curl}u\left(  x\right)  \right\|
^{2}dx\geq C\epsilon^{3/2}\left(  \mu,\nu\right)  \cdot\nu^{-5/2}%
\]
for a suitable universal constant $C>0$. From the previous lemma we have
\[
\underset{k\rightarrow\infty}{\lim\inf}\mu_{n_{k}}\int_{\mathcal{T}}\left\|
D\mathsf{curl}u\left(  x\right)  \right\|  ^{2}dx\geq C\epsilon^{3/2}\left(
\mu,\nu\right)  \cdot\nu^{-5/2}.
\]
Thus the claim of the corollary will follow from the inequality
\begin{align}
\nu\cdot\mu_{n_{k}}\int_{\mathcal{T}}\left\|  D\mathsf{curl}u\left(  x\right)
\right\|  ^{2}dx &  \leq\mu_{n_{k}}\int_{\mathcal{T}}\scal{S_{u}\left
( x\right) \mathsf{curl}u\left( x\right), \mathsf{curl}u\left( x\right
)}\,dx\label{curl balance for Galerkin}\\
&  +\frac{1}{2}\sum_{k\in\Lambda^{(\infty)}}\left|  k\right|  ^{2}\left|
\sigma_{k}\right|  ^{2}.\nonumber
\end{align}
Let us sketch the proof of this inequality (see \cite{FlaCetraro} for more
details). Consider the Galerkin approximations
\[
du^{\left(  n\right)  }+\left[  \nu Au^{\left(  n\right)  }+\pi^{\left(
n\right)  }B\left(  u^{\left(  n\right)  },u^{\left(  n\right)  }\right)
\right]  dt=\sum_{k\in\Lambda^{(n)}}\sigma_{k}\,d\beta_{k}e^{-ik\cdot x}%
\]
described in Appendix B. From It\^{o} formula for $\left\langle Au^{\left(
n\right)  }\left(  t\right)  ,u^{\left(  n\right)  }\left(  t\right)
\right\rangle _{H}$ we get
\begin{align*}
& \left\langle Au^{\left(  n\right)  }\left(  t\right)  ,u^{\left(  n\right)
}\left(  t\right)  \right\rangle _{H}+\int_{0}^{t}2\left\langle Au^{\left(
n\right)  },\nu Au^{\left(  n\right)  }+\pi^{\left(  n\right)  }B\left(
u^{\left(  n\right)  },u^{\left(  n\right)  }\right)  \right\rangle _{H}ds\\
& =\left\langle Au^{\left(  n\right)  }\left(  0\right)  ,u^{\left(  n\right)
}\left(  0\right)  \right\rangle _{H}+M_{t}^{n}+\frac{1}{2}\sum_{k\in
\Lambda_{n}^{(\infty)}}\left|  k\right|  ^{2}\left|  \sigma_{k}\right|  ^{2}%
\end{align*}
where $M_{t}^{n}$ is a square integrable martingale. We have
\[
\left\langle Au^{\left(  n\right)  },\pi^{\left(  n\right)  }B\left(
u^{\left(  n\right)  },u^{\left(  n\right)  }\right)  \right\rangle
_{H}=\left\langle Au^{\left(  n\right)  },B\left(  u^{\left(  n\right)
},u^{\left(  n\right)  }\right)  \right\rangle _{H}%
\]
since $\pi^{\left(  n\right)  }$ is selfadjoint and commutes with $A$. Besides
(\ref{A - curl}) we also have
\[
\left\langle Af,B\left(  g,g\right)  \right\rangle _{H}=\left\langle
\mathsf{curl}f,\left(  g\cdot\nabla\right)  \mathsf{curl}g+S_{g}%
\mathsf{curl}g\right\rangle _{H}%
\]
hence
\[
\left\langle Af,B\left(  f,f\right)  \right\rangle _{H}=\left\langle
\mathsf{curl}f,S_{f}\mathsf{curl}f\right\rangle _{H}%
\]
for every $f,g\in D(A)$. Therefore we have
\begin{align*}
& \left|  \mathsf{curl}u^{\left(  n\right)  }\left(  t\right)  \right|
_{H}^{2}+\int_{0}^{t}\left(  2\nu\left|  D\mathsf{curl}u^{\left(  n\right)
}\right|  _{H}^{2}+\left\langle \mathsf{curl}u^{\left(  n\right)
},S_{u^{\left(  n\right)  }}\mathsf{curl}u^{\left(  n\right)  }\right\rangle
_{H}\right)  ds\\
& \leq\left|  \mathsf{curl}u^{\left(  n\right)  }\left(  0\right)  \right|
_{H}^{2}+M_{t}^{n}+\frac{1}{2}\sum_{k\in\Lambda^{(\infty)}}\left|  k\right|
^{2}\left|  \sigma_{k}\right|  ^{2}.
\end{align*}
This implies (\ref{curl balance for Galerkin}) and the proof is complete.
\end{proof}

\begin{remark}
We cannot conclude (\ref{desiderata}) from the previous corollary without
further (unproved) assumptions on $\mu$ or $\left\{  \mu_{n_{k}}\right\}  $.
This could be just a technical point due to the present lack of better
regularity estimates for the 3D Navier-Stokes equations, or it could be a
facet of a deeper phenomenon. Let us explain it with a cartoon argument. First
recall that it is easy to construct, say on the torus $\mathcal{T}$, a
sequence $\left\{  f_{n}\right\}  $ of functions converging a.s. to zero, but
with $\int_{\mathcal{T}}f_{n}\,dx=1$ (or even $\int_{\mathcal{T}}%
f_{n}\,dx\rightarrow\infty$): just take the mollifiers of a Dirac delta
distribution; if we like, the example can be modified so that $f_{n}$ tend to
develop singularities on a dense zero measure set in $\mathcal{T}$, but the
a.s. limit is still zero. Thus we see that for the limit measure $\mu$ we
could have a small value of $\mu\left[  \int_{\mathcal{T}}\scal{S_{u}%
\left( x\right) \mathsf{curl}u\left( x\right),
\mathsf{curl}u\left( x\right)}\,dx\right]  $ even if some coarse graining
procedure, here represented by the Galerkin approximations, could give us a
large value of $\mu_{n_{k}}\left[  \int_{\mathcal{T}}\scal{S_{u}\left(
x\right) \mathsf{curl}u\left( x\right), \mathsf{curl}u\left( x\right
)}\,dx\right]  $. Such arguments arise the question of the physical meaning of
the true Navier-Stokes equations and possibly of its coarse graining
approximations; this is not our aim, but we wanted to say that the previous
corollary may be considered perhaps as a result of possible physical interest
in itself, even if we cannot rewrite it in the form (\ref{desiderata}). \ 
\end{remark}

\begin{lemma}
Given $\mu\in\mathcal{P}_{{\mathrm{NS}}}^{G}\left(  \nu\right)  $, and every
sequence $\mu_{n_{k}}\in\mathcal{S}^{k_{n}}$ such that $\mu_{k_{n}}$ converges
to $\mu$ in the weak topology of probability measures on $H$, we also have
$\mu_{n_{k}}\rightarrow\mu$ weakly on $\left[  W^{1,3}\left(  \mathcal{T}%
\right)  \right]  ^{3}$.
\end{lemma}

\begin{proof}
From the lemma above, $\left\{  \mu_{n_{k}}\right\}  $ is bounded in
probability on $D(A)$:
\begin{align*}
\mu_{n_{k}}\left(  \left|  Ax\right|  _{H}>R\right)   & =\mu_{n_{k}}\left(
\left|  Ax\right|  _{H}^{2/3}>R^{2/3}\right)  \leq R^{-2/3}\mu_{n}\left(
\left|  A\cdot\right|  _{H}^{2/3}\right) \\
& \leq\frac{C}{R^{2/3}}.
\end{align*}
The embedding of $D(A)$ into $\left[  W^{1,3}\left(  \mathcal{T}\right)
\right]  ^{3}$ is compact:\ recall that Sobolev embedding theorem gives us
$W^{2,2}\subset W^{\beta,\frac{6}{2\beta-1}}$ for every $\beta\in\left(
1,2\right)  $, and the embedding of $W^{\beta,\frac{6}{2\beta-1}}$ in
$W^{1,\frac{6}{2\beta-1}}$ is compact; choose then $\beta=3/2$. Therefore
$\left\{  \mu_{n_{k}}\right\}  $ is tight in $\left[  W^{1,3}\left(
\mathcal{T}\right)  \right]  ^{3}$. Easily we deduce that it converges weakly
to $\mu$ also in $\left[  W^{1,3}\left(  \mathcal{T}\right)  \right]  ^{3}$.
\end{proof}

\begin{corollary}
\label{cor:3d-suff-reg}
If $\mu\in\mathcal{P}_{{\mathrm{NS}}}^{G}\left(  \nu\right)  $ is the weak
limit (in $H$ and thus in $\left[  W^{1,3}\left(  \mathcal{T}\right)  \right]
^{3}$) of a sequence $\mu_{n_{k}}\in\mathcal{S}^{k_{n}}$ such that
\[
\mu_{n_{k}}\left[  \left\|  \cdot\right\|  _{V}^{2+\varepsilon}\right]  \leq C
\]
for some $\varepsilon,C>0$, then
\[
\nu\cdot\mu\int_{\mathcal{T}}\left\|  Du\left(  x\right)  \right\|
^{2}dx=\frac{1}{2}\sum_{k\in\Lambda^{(\infty)}}\left|  \sigma_{k}\right|
^{2}.
\]
If in addition
\[
\mu_{n_{k}}\left[  \left\|  \cdot\right\|  _{V}^{3+\varepsilon}\right]  \leq C
\]
then
\begin{align*}
\nu\cdot\mu\int_{\mathcal{T}}\left\|  D\mathsf{curl}u\left(  x\right)
\right\|  ^{2}dx &  \leq\mu\int_{\mathcal{T}}\scal{S_{u}\left( x\right
) \mathsf{curl}u\left( x\right), \mathsf{curl}u\left( x\right)}\,dx\\
&  +\frac{1}{2}\sum_{k\in\Lambda^{(\infty)}}\left|  k\right|  ^{2}\left|
\sigma_{k}\right|  ^{2}.
\end{align*}
\end{corollary}

\begin{proof}
It is sufficient to apply repeatedly the following fact: if $\mu
_{n}\rightarrow\mu$ weakly in a Polish space $X$, $\varphi\in C\left(
X\right)  $ and $\mu_{n}\left[  \left|  \varphi\right|  ^{1+\varepsilon
}\right]  \leq C$, then $\mu_{n}\left[  \varphi\right]  \rightarrow\mu\left[
\varphi\right]  $. This fact is well know but we provide the proof for
completeness. Let $Y_{n}$ and $Y$ be r.v.'s with law $\mu_{n}$ and $\mu$
resp., with values in $X$, such that $Y_{n}\rightarrow Y$ a.s. in $X$. Then
$\mu_{n}\left[  \varphi\right]  =E\left[  \varphi\left(  Y_{n}\right)
\right]  $, $\mu\left[  \varphi\right]  =E\left[  \varphi\left(  Y\right)
\right]  $, so by Vitali convergence theorem it is sufficient to prove that
$\varphi\left(  Y_{n}\right)  $ is uniformly integrable. We have
\[
E\left[  \varphi\left(  Y_{n}\right)  1_{\varphi\left(  Y_{n}\right)
\geq\lambda}\right]  \leq\left(  E\left[  \varphi\left(  Y_{n}\right)
^{p}\right]  \right)  ^{1/p}P\left(  \varphi\left(  Y_{n}\right)  \geq
\lambda\right)  ^{1/q}\leq C\lambda^{-\delta}.
\]
Thus the uniform integrability is proved and the proof is complete.
\end{proof}

\begin{corollary}
\label{stretching under additional hp}Let $\mathcal{M}\subset\mathcal{P}%
\times\mathbb{R}_{+}$, with $\mathcal{M}_{\nu}\subset\mathcal{P}%
_{{\mathrm{NS}}}^{G}\left(  \nu\right)  $, be a family with the K41 scaling
law, in the sense of definition \ref{defk41}. Assume that every $\mu$ in
$\mathcal{M}$ is the weak limit of a sequence $\mu_{n_{k}}\in\mathcal{S}%
^{k_{n}}$ such that
\[
\mu_{n_{k}}\left[  \left\|  \cdot\right\|  _{V}^{3+\varepsilon}\right]  \leq C
\]
for some $\varepsilon,C>0$. Then there exists $\nu_{0}>0$ and $C>0$ such that
(\ref{desiderata}) holds for every $\nu\in\left(  0,\nu_{0}\right)  $ and
every $\mu\in\mathcal{M}_{\nu}$.
\end{corollary}

\begin{remark}
If K41 scaling law holds then vortex stretching must be intense.
Heuristically, no geometrical depletion of such stretching may occur (in
contrast to the 2D case where the stretching term is zero because
$\mathsf{curl}u\left(  x\right)  $ is aligned with the eigenvector of
eigenvalue zero of $S_{u}\left(  x\right)  $): indeed, if we extrapolate the
behaviour $E\left[  \left|  Du\right|  ^{2}\right]  \sim\frac{1}{\nu}$ as
$Du\sim\frac{1}{\sqrt{\nu}}$, $\mathsf{curl}u\sim\frac{1}{\sqrt{\nu}}$, then
we get $E\left[  S_{u}\mathsf{curl}u\cdot\mathsf{curl}u\right]  \sim\frac
{1}{\nu\sqrt{\nu}} $ if there is no help from the geometry. Another way to
explain this idea is the following sort of generalised H\"{o}lder inequality.
\end{remark}

\begin{corollary}
\label{3Dsecondcorollary}Let $\mathcal{M}\subset\mathcal{P}\times
\mathbb{R}_{+}$, with $\mathcal{M}_{\nu}\subset\mathcal{P}_{{\mathrm{NS}}}%
^{G}\left(  \nu\right)  $, be a family with the K41 scaling law, fulfilling
the assumptions of corollary \ref{stretching under additional hp}. Then there
exists $\nu_{0}>0$ and $C>0$ such that
\[
\left(  \mu\int_{\mathcal{T}}\left\|  Du\right\|  ^{2}dx\right)  ^{1/2}\leq
C\left(  \mu\left[  \int_{\mathcal{T}}\left\|  S_{u}\mathsf{curl}%
u\cdot\mathsf{curl}u\right\|  ^{2}dx\right]  \right)  ^{1/3}%
\]
for every $\nu\in\left(  0,\nu_{0}\right)  $ and every $\mu\in\mathcal{M}%
_{\nu}$.
\end{corollary}

\begin{proof}
From the previous corollary and the definition of $\epsilon\left(  \mu
,\nu\right)  $ we have
\begin{align*}
\left(  \mu\left[  \int_{\mathcal{T}}\left\|  S_{u}\mathsf{curl}%
u\cdot\mathsf{curl}u\right\|  ^{2}dx\right]  \right)  ^{1/3} &  \geq\left(
C\epsilon^{3/2}\left(  \mu,\nu\right)  \cdot\nu^{-3/2}\right)  ^{1/3}\\
&  =C^{\prime}\epsilon^{1/2}\left(  \mu,\nu\right)  \cdot\nu^{-1/2}\\
&  =C^{\prime}\left(  \mu\int_{\mathcal{T}}\left\|  Du\right\|  ^{2}dx\right)
^{1/2}.
\end{align*}
The proof is complete.
\end{proof}

\section{Necessary and sufficient conditions for K41}
\label{sec:NecSuf}

We continue with the notations and concepts just introduced in the last
section on the 3D case.

The result of this section can be formulated for definition \ref{defk41}, but
the presence of the factor $\epsilon\left(  \mu,\nu\right)  ^{-1/4}$ in the
definition of $\eta\left(  \mu,\nu\right)  $ makes some statements much less
direct. So, having in mind the exploratory character of these equivalent
conditions, we prefer to adopt a simplified form of our definition of the K41 scaling law.

\begin{definition}
\label{defk41bis}We say that a scaling law of K41 type holds true for a set
$\mathcal{M}\subset\mathcal{P}\times\mathbb{R}_{+}$ if there exist $\nu_{0}%
>0$, $C>c>0$, $C_{0}>0$, and a monotone function $R_{0}\colon(0,\nu
_{0}]\rightarrow\mathbb{R}_{+}$ with $R_{0}\left(  \nu\right)  >C_{0}$ and
$\lim_{\nu\rightarrow0}R_{0}\left(  \nu\right)  =+\infty$, such that the
bound
\begin{equation}
c\cdot r^{2/3}\leq S_{2}^{\mu}\left(  r\right)  \leq C\cdot r^{2/3}%
\label{e:K41}%
\end{equation}
holds for every pair $(\mu,\nu)\in\mathcal{M}$ and every $r$ such that $\nu
\in(0,\nu_{0}]$ and
\[
C_{0}\nu^{3/4}<r<\nu^{3/4}R_{0}\left(  \nu\right)  .
\]
\end{definition}

Recalling that $\eta\left(  \mu,\nu\right)  =\nu^{3/4}\epsilon\left(  \mu
,\nu\right)  ^{-1/4}$, we see that this definition is equivalent to
\ref{defk41} if there exist $\epsilon_{1}>\epsilon_{0}>0$ such that
\[
\epsilon_{0}\leq\epsilon\left(  \mu,\nu\right)  \leq\epsilon_{1}%
\]
for all $(\mu,\nu)\in\mathcal{M}$. Unfortunately, in 3D only the upper
bound can be proven. However, this could be just a technical problem
due to the fact that we can only use weak solutions (for slightly more regular
solutions Corollary~\ref{cor:3d-suff-reg} implies that
$\epsilon\left(  \mu,\nu\right)  $ would be bounded from above and below).

Consider the auxiliary stochastic Navier-Stokes equations
\begin{equation}
\frac{\partial\widetilde{u}}{\partial t}(t,x)+\left(
\widetilde{u}\left(  t,x\right)  \cdot\nabla\right)  \widetilde{u}\left(
t,x\right)  +\nabla\widetilde{p}\left(  t,x\right)
=\tilde{\nu}\triangle\widetilde{u}\left(  t,x\right)  +\sum_{
k\in\Lambda^{(\infty)}_L}\sigma_k{\dot\beta}_k(t)
e^{-ik\cdot x} \label{e:SNS}%
\end{equation}%
on the torus $\left[  0,L\right]  ^{3}$ with $\Div \widetilde{u}=0$ and
periodic boundary conditions (the set $\Lambda^{(\infty)}_L$ is defined in
(\ref{LambdaL})). As we shall see below (see the next section and
lemma \ref{appendixlemmascaling}), we obtain this equation when we perform the
following scaling transformation on the solutions $u$ of the original equation
(\ref{stochNS}):
\[
\widetilde{u}\left(  t,x\right)  =L^{1/3}u(L^{-2/3}t,L^{-1}x)
\]
(and a suitably defined $\widetilde{p}\left(  t,x\right)  $). The value of
$\tilde{\nu}$ under this transformation is
\[
\tilde{\nu}=\nu L^{4/3}.
\]
This scaling transformation has been introduced in the mathematical-physics
literature, see~\cite{Kup}. What makes it special is that no coefficient
depending on the scale parameter appears in front of the noise, so the energy
input per unit of time and space is the same for every $L$. Heuristically, if we
believe in a cascade picture of the energy (without essential inverse
cascade), this invariance of the energy input should imply that the small
scale properties of (\ref{stochNS}) and (\ref{e:SNS}) are the same, namely
that they are invariant under this transformation; this should lead to the K41
scaling law.

Similarly to the case $L=1$, we may introduce the (non empty) set
$\mathcal{P}_{{\mathrm{NS}}}^{G}\left(  \tilde{\nu},L\right)  $ of limit
points of the (homogeneous and isotropic) invariant measures of the
corresponding Galerkin approximations.

Let us denote by $\mathcal{P}_{{\mathrm{NS}}}^{G}$ the set of \textit{all
pairs} $(\mu,\nu)$ such that $\mu\in\mathcal{P}_{{\mathrm{NS}}}^{G}\left(
\nu\right)  $. Similarly, let us denote by $\tilde{\mathcal{P}}_{{\mathrm{NS}%
}}^{G}$ the set of \textit{all triples} $(\mu,\tilde{\nu},L)$ such that
$\mu\in\mathcal{P}_{{\mathrm{NS}}}^{G}\left(  \tilde{\nu},L\right)  $.

\subsection{Basic equivalent condition}

The following condition seems interesting since it looks rather qualitative,
in contrast to Definition~\ref{defk41bis}, and shows that the exponent $2/3$
arises from the scaling properties of the stochastic Navier-Stokes equations.

Let us introduce the notation $\mathcal{P}_{L}$ for the set of probability
measures analogous to $\mathcal{P}$, but on the torus $\left[  0,L\right]
^{3} $. Denote by $\mathcal{P}_{\cdot}\times\mathbb{R}_{+}^{2}$ the set of all
triples $\left(  \mu,\tilde{\nu},L\right)  $ such that $\left(  \tilde{\nu
},L\right)  \in\mathbb{R}_{+}^{2}$ and $\mu\in\mathcal{P}_{L}$. In the next
definition and later on we use the notation $\mu\left[  \left\|  u\left(
e\right)  -u(0)\right\|  ^{2}\right]  $ when $\mu\in\mathcal{P}_{L}$ (and
other similar mean values): this means
\[
\mu\left[  \left\|  u\left(  e\right)  -u(0)\right\|  ^{2}\right]
=\int_{H_{L}}\left\|  u\left(  e\right)  -u(0)\right\|  ^{2}d\mu\left(
u\right)\;,
\]
where $H_{L}$ has been introduced in section \ref{section 1.1}.

\begin{definition}
We call admissible region a set $D\subset\mathbb{R}_{+}^{2}$ of the following
form:
\[
D=\left\{  \left(  \tilde{\nu},L\right)  \in\mathbb{R}_{+}^{2};\tilde{\nu}%
\in\left(  0,\nu_{0}\right)  ,L>\tilde{R}_{0}\left(  \tilde{\nu}\right)
\right\}
\]
where $\tilde{\nu}_{0}>0$ and $\tilde{R}_{0}\colon(0,\tilde{\nu}%
_{0}]\rightarrow\lbrack1,\infty)$ is a strictly decreasing function with
$\tilde{R}_{0}(\tilde{\nu})\rightarrow\infty$ as $\tilde{\nu}\rightarrow0$. 
\end{definition}

An
admissible region is depicted in the left-hand side of Figure~\ref{fig:effectK} below.

\begin{condition}
\label{condA} A subset $\tilde{\mathcal{M}}\subset\mathcal{P}_{\cdot}%
\times\mathbb{R}_{+}^{2}$ is said to satisfy Condition~\ref{condA} if there
exist an admissible region $D\subset\mathbb{R}_{+}^{2}$ and two constants
$C>c>0$ such that
\begin{equation}
c\leq\mu\left[  \left\|  u\left(  e\right)  -u(0)\right\|  ^{2}\right]  \leq
C\label{e:condA}%
\end{equation}
for every $(\mu,\tilde{\nu},L)\in\tilde{\mathcal{M}}$ with $\left(  \tilde
{\nu},L\right)  \in D$.
\end{condition}

\begin{proposition}
\label{mainprop} The set $\tilde{\mathcal{P}}_{{\mathrm{NS}}}^{G}$ satisfies
Condition~\ref{condA} if and only if the set $\mathcal{P}_{{\mathrm{NS}}}^{G}$ has a
scaling law of K41 type, in the sense of Definition~\ref{defk41bis}.
\end{proposition}

\begin{proof}
Given $R>0$, consider the mapping $S_{R}\colon\mathcal{H}_{R}\rightarrow
\mathcal{H}$ defined by
\begin{equation}
(S_{R}u)(x)=R^{1/3}u(Rx)\;.
\end{equation}
This mapping induces a mapping ${\CS}$ from $\mathcal{P}\times\mathbb{R}_{+}^{2}$
to $\mathcal{P}\times\mathbb{R}_{+}$ by
\begin{equation}
{\CS}(\mu,\tilde{\nu},\tilde{r})=\bigl(S_{\tilde{r}}^{\ast}\mu,\tilde{\nu}%
\tilde{r}^{-4/3}\bigr)\;.
\end{equation}
It follows immediately from Theorem \ref{appendixtheoremscaling} that one has
\begin{equation}
\mathcal{P}_{{\mathrm{NS}}}^{G}={\CS}(\tilde{\mathcal{P}}_{{\mathrm{NS}}}%
^{G})\;,\quad\hbox{and}\quad\tilde{\mathcal{P}}_{{\mathrm{NS}}}^{G}={\CS}%
^{-1}(\mathcal{P}_{{\mathrm{NS}}}^{G})\;.
\end{equation}
Furthermore, it follows immediately from the above definitions that if
$(\mu,\nu)={\CS}(\tilde{\mu},\tilde{\nu},\tilde{r})$, then
\begin{equation}
S_{2}^{\mu}(r)=r^{2/3}\int_{H_{\tilde{r}}}\left\|  u\left(  e\right)
-u(0)\right\|  ^{2}d\tilde{\mu}\left(  u\right)  \;.
\end{equation}
It therefore follows that, in order to prove the equivalence between
Condition~\ref{condA} and K41, it suffices to show that the domains of validity of eq.~\ref{e:condA}
and of eq.~\ref{e:K41} are the same (with possibly different constants and
functions $R_{0}$ and $\tilde{R}_{0}$), provided that $(\nu,r)$ and
$(\tilde{\nu},\tilde{r})$ are related by
\begin{equation}
\tilde{\nu}=\nu r^{-4/3}\;,\qquad\tilde{r}=r^{-1}\;.\label{e:relnur}%
\end{equation}
We denote by $K\colon(\nu,r)\mapsto(\tilde{\nu},\tilde{r})$ the above map.

\medskip
\noindent\textbf{Condition~\ref{condA} implies K41.} The domain of validity of
eq.~\ref{e:condA} is given by
\begin{equation}
\tilde{\nu}\leq\tilde{\nu}_{0}\;,\qquad\tilde{r}\geq\tilde{R}_{0}(\tilde{\nu
})\;.
\end{equation}
Under the map $K^{-1}$, this becomes
\begin{equation}
r\geq\Bigl({\frac{\nu}{\tilde{\nu}_{0}}}\Bigr)^{3/4}\equiv C_{0}\nu
^{3/4}\;,\qquad{\frac{1}{r}}\geq\tilde{R}_{0}(\nu r^{-4/3}%
)\;.\label{e:imcondA}%
\end{equation}
Both domains are shown in Fig.~\ref{fig:effectK}.

\begin{figure}
\begin{center}
\mhpastefig[3/2]{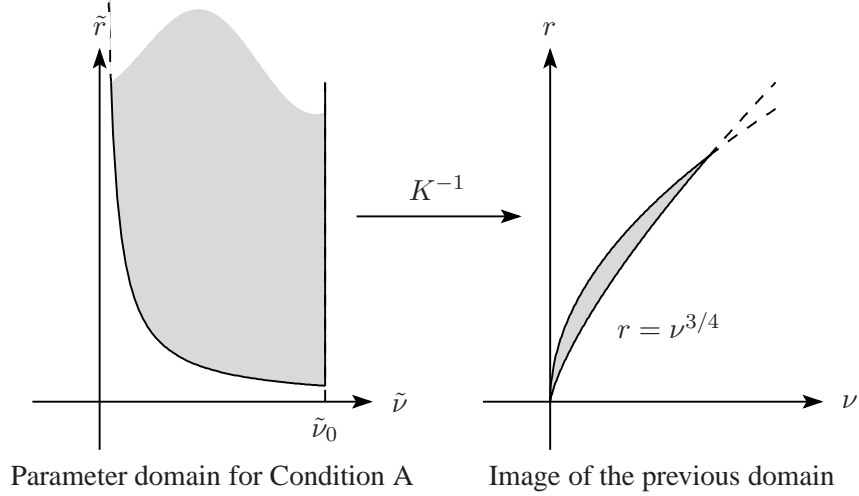}
\end{center}
\caption{Effect of $K^{-1}$ on an admissible domain}
\label{fig:effectK}
\end{figure}

Defining the strictly decreasing function $F(x)=x^{-3/4}\tilde{R}_{0}(x)$, the
second condition of eq.~\ref{e:imcondA} is of course  equivalent to
\begin{equation}
\nu^{-3/4}\geq F(\nu r^{-4/3})\;.\label{e:equ1}%
\end{equation}
This condition (as can be inferred from the Fig.~\ref{fig:effectK}), can only be
satisfied simultaneously with the first condition in eq.~\ref{e:imcondA} if
$\nu\leq\nu_{0}\equiv F(\tilde{\nu}_{0})^{-4/3}$. On $(0,\nu_{0}]$ this
domain, eq.~\ref{e:equ1} is equivalent to
\begin{equation}
r\leq\Bigl({\frac{\nu}{{F^{-1}(\nu^{-3/4})}}}\Bigr)^{3/4}\equiv\nu^{3/4}%
R_{0}(\nu)\;,
\end{equation}
where $R_{0}(x)=\bigl(F^{-1}(x^{-3/4})\bigr)^{-3/4}$. Additionally $R_{0}$ is
well-defined on $(0,\nu_{0}]$ and that it is greater than $C_{0}$ on this
domain. Furthermore, since $F$ is decreasing, $R_{0}$ is strictly decreasing
and it is easy to check that $\lim_{x\rightarrow0}R_{0}(x)=\infty$ because the
same property holds for $F$.

\medskip
\noindent\textbf{K41 implies Condition~\ref{condA}.} The domain of validity of K41 is
given by
\begin{equation}\label{e:domK41}
\nu\leq\nu_{0}\;,\qquad r\nu^{-3/4}\in\lbrack C_{0},R_{0}(\nu)]\;.
\end{equation}
Under the map $K$, this becomes
\begin{equation}
\tilde{\nu}\tilde{r}^{-4/3}\leq\nu_{0}\;,\qquad\tilde{\nu}^{-3/4}\in\lbrack
C_{0},R_{0}(\tilde{\nu}\tilde{r}^{-4/3})]\;.\label{e:imcondK41}%
\end{equation}
The second condition can be rewritten as
\begin{equation}
\tilde{\nu}\in\lbrack G(\tilde{\nu}\tilde{r}^{-4/3}),\tilde{\nu}_{0}]\;,
\end{equation}
where we defined $\tilde{\nu}_{0}=C_{0}^{-4/3}$ and $G(x)=R_{0}(x)^{-4/3}$.
Both of these domains are shown in Figure~\ref{fig:effectKinv}.

\begin{figure}
\begin{center}
\mhpastefig[3/2]{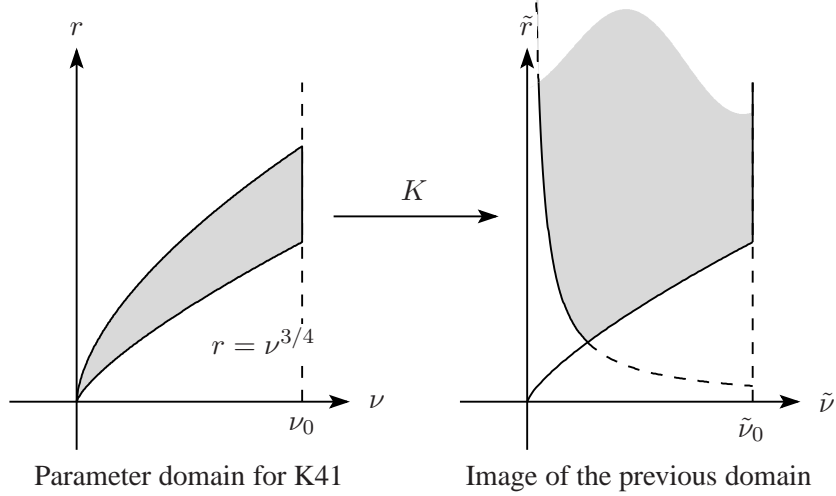}
\end{center}
\caption{Effect of $K$ on a domain of the type (\ref{e:domK41})}
\label{fig:effectKinv}
\end{figure}

We can rewrite as above the condition $\tilde{\nu}\geq G(\tilde{\nu}\tilde
{r}^{-4/3})$ as
\begin{equation}
\tilde{r}\geq\Bigl({\frac{\tilde{\nu}}{G^{-1}(\tilde{\nu})}}\Bigr)^{3/4}%
\equiv\tilde{R}_{0}(\tilde{\nu})\;.
\end{equation}
Again, it is an easy exercise to show that $\tilde{R}_{0}$ as defined above is
monotone and satisfies $\lim_{x\rightarrow0}\tilde{R}_{0}(x)=\infty$. The only
points that remain to be clarified are:
\begin{itemize}
\item[a.]We haven't taken the first equation in eq.~\ref{e:imcondK41} into account.
\item[b.] The domain of definition of $R_{0}$ may not extend to $\tilde{\nu
}_{0}$.
\end{itemize}
Both problems can be solved at once by simply choosing a smaller value for
$\tilde{\nu}_{0}$
\end{proof}

%

\subsection{Necessary and sufficient conditions in terms of high and low modes}

Although Condition~\ref{condA} contains only bounds (at finite distance points) and not
scaling exponents (with small distant points), and thus in principle it
represents a progress in the direction of analysis of K41, it still looks
difficult to verify or disprove it for Navier-Stokes equations, since it is
rather unusual to work with the difference of a solution at two points. This
is the main motivation for the following new necessary and sufficient conditions.

Looking at them on the other direction, as \textit{necessary} conditions for
K41, they declare that under K41 the energy of high modes is bounded and the
enstrophy of low modes is bounded, an information with a certain physical content.

In this section, for notational simplicity, we drop the tildes in our notation. 
Recall that an admissible region is defined by
\[
D=\left\{  \left(  \nu,L\right)  \in\mathbb{R}_{+}^{2};\nu\in\left(  0,\nu
_{0}\right)  ,L>R_{0}\left(  \nu\right)  \right\}  \;,
\]
and that Condition~\ref{condA} requires
\[
c\leq\mu\left[  \left\|  u\left(  e\right)  -u(0)\right\|  ^{2}\right]  \leq C
\]
for every $(\mu,\nu,L)$ with $\left(  \nu,L\right)  \in D$. 

We start with a preparatory lemma which depends on the scaling properties of stochastic Navier-Stokes equation in an essential way. This is the only point in this section where specific informations about the measures are used.

\begin{lemma}
\label{lemmalambda}If $\tilde{\mathcal{P}}_{{\mathrm{NS}}}^{G}$ satisfies
Condition~$A$ then there exist constants $C^{\prime}>c^{\prime}>0$ and an
admissible region $D^{\prime}$ such that
\[
c^{\prime}\leq\sum_{e}\int_{\frac{1}{2}}^{\frac{3}{2}}\mu\left[  \left\|
u\left(  \lambda e\right)  -u(0)\right\|  ^{2}\right]  d\lambda\leq C^{\prime}%
\]
for every $(\mu,\nu,L)\in\tilde{\mathcal{P}}_{{\mathrm{NS}}}^{G}$ with
$(\nu,L)\in D^{\prime}$. The sum $\sum_{e}$ is extended to all coordinate
unitary vectors. We simply have $C^{\prime}=\left(  1.5^{2/3}d\right)  \cdot
C$, $c^{\prime}=\left(  0.5^{2/3}d\right)  \cdot c$, $D^{\prime}$ defined by
$0.5^{4/3}\cdot\nu_{0}$ and $1.5R_{0}\left(  1.5^{-4/3}\nu\right)  $, where
$\nu_{0}$ and $R_{0}\left(  \nu\right)  $ define $D$.
\end{lemma}

\begin{proof}
Given $\lambda\in\left[  \frac{1}{2},\frac{3}{2}\right]  $ and $(\mu,\nu
,L)\in\tilde{\mathcal{P}}_{{\mathrm{NS}}}^{G}$, namely $\mu\in\mathcal{P}%
_{NS}^{G}\left(  \nu,L\right)  $, consider the measure $\mu_{\lambda}$ that
corresponds to $\mu$ under the transformation $u\mapsto\lambda^{-1/3}%
u(\lambda.)$ used in the previous section, having the property
\[
\mu\left[  \left\|  u\left(  \lambda e\right)  -u(0)\right\|  ^{2}\right]
=\lambda^{2/3}\mu_{\lambda}\left[  \left\|  u\left(  e\right)  -u(0)\right\|
^{2}\right]  .
\]
By Theorem~\ref{appendixtheoremscaling} we know that $\mu_{\lambda}%
\in\mathcal{P}_{NS}^{G}\left(  \nu\lambda^{-4/3},L/\lambda\right)  $, hence
$(\mu_{\lambda},\nu\lambda^{-4/3},L/\lambda)\in\tilde{\mathcal{P}%
}_{{\mathrm{NS}}}^{G}$. Thus Condition~\ref{condA} implies
\[
c\leq\mu_{\lambda}\left[  \left\|  u\left(  e\right)  -u(0)\right\|
^{2}\right]  \leq C
\]
if $\nu\lambda^{-4/3}<\nu_{0}$ and $L/\lambda>R_{0}\left(  \nu\lambda
^{-4/3}\right)  $. The first condition is true if $\nu<0.5^{4/3}\nu_{0}$. The
second one if $L>1.5R_{0}\left(  1.5^{-4/3}\nu\right)  $. The proof can now be
easily completed.
\end{proof}

Let us use some Fourier analysis on the torus $T_{L}=\left[  0,L\right]  ^{d}%
$ (see also Appendix \ref{appendixtwo}). Every $u\in H_{L}$ is given by
\[
u\left(  x\right)  =\sum_{k\in\Lambda^{\left(  \infty\right)_L  }%
}e^{-ik\cdot x}\widehat{u}\left(  k\right)
\]
where
\[
\widehat{u}\left(  k\right)  :=L^{-3}\int_{T_{L}}e^{ik\cdot x}u\left(
x\right)  dx
\]
and we have Parseval identity
\[
L^{-3}\int_{T_{L}}\left\|  u\left(  x\right)  \right\|  ^{2}dx=\sum
_{k\in\Lambda^{\left(  \infty\right)}_L}\left\|  \widehat{u}\left(
k\right)  \right\|  ^{2}.
\]

We introduce another condition which requires  the sum of the enstrophy of low modes and energy of high modes to be finite and bounded away from zero. 

\begin{condition}
\label{condB} A subset $\tilde{\mathcal{M}}\subset\mathcal{P}_{\cdot}%
\times\mathbb{R}_{+}^{2}$ is said to satisfy Condition~\ref{condB} if there
exist an admissible region $D\subset\mathbb{R}_{+}^{2}$ and two constants
$C>c>0$ such that
\[
c\leq\sum_{\substack{k\in\Lambda_{L}^{\left(  \infty\right)  }\\\left\|
k\right\|  \leq 1}}\|k\|^2\mu\left[  \left\|  \widehat
{u}\left(  k\right)  \right\|  ^{2}\right]  +\sum_{\substack{k\in\Lambda
_{L}^{\left(  \infty\right)  }\\\left\|  k\right\|  >1}}\mu\left[  \left\|
\widehat{u}\left(  k\right)  \right\|  ^{2}\right]  \leq C
\]
for every $(\mu,\nu,L)\in\tilde{\mathcal{P}}_{{\mathrm{NS}}}^{G}$ such that
$(\nu,L)\in D$. 
\end{condition}

With this definition, we may establish a first basic theorem as a corollary of
the previous lemma.

\begin{theorem} 
Condition~\ref{condA}
implies Condition~\ref{condB}.
\end{theorem}

\begin{remark}
We understand that constants and admissible regions involved
in Conditions~\ref{condA} and~\ref{condB} are not necessarily the same.
\end{remark}

\begin{proof}
For every $u\in H_{L}$ we have
\begin{align*}
\left\|  u\left(  \lambda e\right)  -u(0)\right\|  ^{2}  & =L^{-3}\int_{T_{L}%
}\left\|  u\left(  x+\lambda e\right)  -u\left(  x\right)  \right\|  ^{2}dx\\
& =\sum_{k\in\Lambda^{(\infty)}_L}\left|  e^{ik\cdot\lambda e
}-1\right|  ^{2}\left\|  \widehat{u}\left(  k\right)  \right\|  ^{2}%
\end{align*}
and thus, for every $\mu\in\mathcal{P}_{NS}^{G}\left(  \nu,L\right)  $ we
have
\begin{align*}
& \sum_{e}\int_{\frac{1}{2}}^{\frac{3}{2}}\mu\left[  \left\|  u\left(  \lambda
e\right)  -u(0)\right\|  ^{2}\right]  d\lambda\\
& =\sum_{k\in\Lambda^{(\infty)}_L}\left(  \sum_{e}\int
_{\frac{1}{2}}^{\frac{3}{2}}\left|  e^{ik\cdot\lambda e}-1\right|
^{2}d\lambda\right)  \mu\left[  \left\|  \widehat{u}\left(  k\right)
\right\|  ^{2}\right]  .
\end{align*}
But there exist universal constants $C^{\prime}>c^{\prime}>0$ such that
\[
c^{\prime}(\left\|k\right\|^{2}\wedge1)\leq\sum_{e}%
\int_{\frac{1}{2}}^{\frac{3}{2}}\left|  e^{ik\cdot\lambda e}-1\right|
^{2}d\lambda\leq C^{\prime}(\left\|k\right\|^{2}\wedge1).
\]
Therefore, the quantities
\[
\sum_{e}\int_{\frac{1}{2}}^{\frac{3}{2}}\mu\left[  \left\|  u\left(  \lambda
e\right)  -u(0)\right\|  ^{2}\right]  d\lambda
\]
and
\[
\sum_{k\in\Lambda^{(\infty)}_L}\left(\left\|k\right\|^{2}
\wedge1\right)  \mu\left[  \left\|  \widehat{u}\left(  k\right)  \right\|
^{2}\right]
\]
are ``equivalent'', up to universal constants. This proves the claim.
\end{proof}

We have at least a partial converse of the previous result if we require that in the admissible region the enstrophy of high modes is by itself bounded away from zero. Then we introduce the following condition:

\begin{condition}
\label{condC} A subset $\tilde{\mathcal{M}}\subset\mathcal{P}_{\cdot}%
\times\mathbb{R}_{+}^{2}$ is said to satisfy Condition~\ref{condC} if there
exist an admissible region $D\subset\mathbb{R}_{+}^{2}$ and two constants
$C>c>0$ such that
\[
c\leq
\sum_{\substack{k\in\Lambda_{L}^{\left(  \infty\right)  }\\\left\|
k\right\|  \leq 1/2}}\|k\|^2\mu\left[  \left\|  \widehat
{u}\left(  k\right)  \right\|  ^{2}\right]
\leq\sum_{\substack{k\in\Lambda_{L}^{\left(  \infty\right)  }\\\left\|
k\right\|  \leq 1}}\|k\|^2\mu\left[  \left\|  \widehat
{u}\left(  k\right)  \right\|  ^{2}\right]  +\sum_{\substack{k\in\Lambda
_{L}^{\left(  \infty\right)  }\\\left\|  k\right\|  >1}}\mu\left[  \left\|
\widehat{u}\left(  k\right)  \right\|  ^{2}\right]  \leq C
\]
for every $(\mu,\nu,L)\in\tilde{\mathcal{P}}_{{\mathrm{NS}}}^{G}$ such that
$(\nu,L)\in D$. 
\end{condition}

Note that Condition~\ref{condC} implies directly Condition~\ref{condB}. What is more interesting is the following:

\begin{proposition}
Condition~\ref{condC} implies Condition~\ref{condA}.
\end{proposition}

\begin{proof}
We have
\[
\sum_{e}\left|  e^{ik\cdot e}-1\right|  ^{2}\leq C\left(
\left\|k\right\|^{2}\wedge1\right).
\]
for every $k$.
Moreover if $\|k\| \le 1/2$ we have
\[
c\|k\|^{2} \le \sum_{e}\left|  e^{ik\cdot e}-1\right|  ^{2}
\]
for some constant $c> 0$.
The claim then follows from the next lemma and the following inequality
\begin{align*}
\sum_{k\in\Lambda_{L}^{\left(  \infty\right) } }\left(  \sum_{e}\left|
e^{ik\cdot e}-1\right|  ^{2}\right)  \mu\left[  \left\|  \widehat
{u}\left(  k\right)  \right\|  ^{2}\right]
&\ge \sum_{\substack{k\in\Lambda_{L}^{\left(  \infty\right)  }\\\left\|
k\right\|  \leq 1/2}}\left(  \sum_{e}\left|
e^{ik\cdot e}-1\right|  ^{2}\right)  \mu\left[  \left\|  \widehat
{u}\left(  k\right)  \right\|  ^{2}\right]  
\\ &\ge c \sum_{\substack{k\in\Lambda_{L}^{\left(  \infty\right)  }\\\left\|
k\right\|  \leq 1/2}}\|k\|^{2}  \mu\left[  \left\|  \widehat
{u}\left(  k\right)  \right\|  ^{2}\right]  .
\end{align*}
\end{proof}

\begin{lemma}
\bigskip$\tilde{\mathcal{P}}_{{\mathrm{NS}}}^{G}$ satisfies Condition~$A$ if
and only if it satisfies the following \textbf{Condition }$\mathbf{A}^{\prime
}$: there exist $C>c>0$, and an admissible region $D$ such that
\[
c\leq\sum_{k\in\Lambda^{(\infty)}_L}\left(  \sum_{e}\left|
e^{ik\cdot e}-1\right|  ^{2}\right)  \mu\left[  \left\|  \widehat
{u}\left(  k\right)  \right\|  ^{2}\right]  \leq C
\]
for every $(\mu,\nu,L)\in\tilde{\mathcal{P}}_{{\mathrm{NS}}}^{G}$ such that
$(\nu,L)\in D$.
\end{lemma}

\begin{proof}
From previous computations, we know that for every $\mu\in\mathcal{P}_{NS}%
^{G}\left(  \nu,L\right)  $ we have
\begin{align*}
\sum_{e}\mu\left[  \left\|  u\left(  e\right)  -u(0)\right\|  ^{2}\right]
=\sum_{k\in\Lambda^{\left(  \infty\right)  }_L}\left(  \sum_{e}\left|
e^{ik\cdot e}-1\right|  ^{2}\right)  \mu\left[  \left\|  \widehat
{u}\left(  k\right)  \right\|  ^{2}\right]  .
\end{align*}

This proves the claim.
\end{proof}
%


\section{A random eddy model}
\label{sec:Eddy}

We now exhibit a model having the property stated in the conjecture, and other
heuristically meaningful properties for a turbulent velocity field. The model
is mathematically rigorous but it is~\textit{not} derived from the
Navier-Stokes equations, it is just a cartoon of what we believe to resemble the
turbulent 3D field given by the Navier-Stokes equations. Therefore the only
merit of the following result is to show that there exists a field with the
property stated in the conjecture, and such a field is not just an artificial
example but it is strongly inspired by numerical and physical observations of
turbulent fluids.

\medskip For simplicity we work in the full three-dimensional space
$\mathbb{R}^{3}$, instead of the torus $\mathcal{T}$.

The model should be thought of as a random collection of \emph{vortex filaments},
i.e. concentrations of vorticity around one-dimensional continuous curves. The
filaments will be of various kind, from very elongated ones, whose existence
is well documented in numerical observations of fully developed turbulence, to
other more ``eddy-like'' and symmetric.

The basic ingredient of the construction is a vortex filament of length $T$,
thickness $\ell$ and core velocity $U$, which is stochastically modelled
around a ``Brownian'' core: consider a 3d-Brownian motion $\{X_{t}%
\}_{t\in\lbrack0,T]}$ starting from a point $X_{0}$. This is the backbone of
the vortex filament whose vorticity field is given by
\begin{equation}
\xi_{\text{single}}(x)=\frac{U}{\ell^{2}}\int_{0}^{T}\rho_{\ell}(x-X_{t})
\circ dX_{t}.
\end{equation}
where $\circ dX$ denote Stratonovich integration. The letter $t$, that
sometimes we shall also call time, is not physical time but just the parameter
of the curve. We assume that $\rho_{\ell}(x)=\rho(x/\ell)$ for a radially
symmetric measurable bounded (smooth) function $\rho$ with compact support in
the ball $B(0,1)$ (the unit ball in 3d Euclidean space). Heuristically
$\xi_{\text{single}}(x)$ is an average of the ``directions'' $dX_{t}$ for
points $X_{t}$ in the ball $B(x,\ell)$. The various parameters $U,\ell,T$ have
to be thought of as giving the ``typical'' magnitudes of the respective
properties. It should be noted that $\xi$ is not a ``real'' vorticity field (since in this model its divergence is not zero) but should be understood as providing the contribution to the fuild vorticity coming from the eddies.

The velocity field $u$ is generated from $\xi$ according to the Biot-Savart
relation
\begin{equation}
u_{\text{single}}(x)=\frac{U}{\ell^{2}}\int_{0}^{T}\mathcal{K}_{\ell}(x-X_{t}%
)\wedge\circ dX_{t}%
\end{equation}
where the vector kernel $\mathcal{K}_{\ell}(x)$ is defined as
\begin{equation}
\mathcal{K}_{\ell}(x)=\frac{1}{4\pi}\int_{B(0,\ell)}\rho_{\ell}(y)\frac{x-y}{|x-y|^{3}%
}dy.
\end{equation}
We want to describe a random superposition of infinitely many independent
Brownian vortex filaments, uniformly distributed in space, each of which will
be associated with intensity-thickness-length parameters $(U,\ell,T)$
``randomly drawn'' according to a measure $\gamma$. The total vorticity of the
fluid is the sum of the vorticities of the single filaments, so, by linearity of
the relation vorticity-velocity, the total velocity field will be the sum of
the velocity fields of the single filaments.

The correct mathematical implementation of this heuristic picture is given by
the construction of a Poisson random measure on a suitable space.

Let $\Xi$ be the metric space
\[
\Xi=\{(U,\ell,T,X)\in\mathbb{R}_{+}^{3}\times C([0,1];\mathbb{R}^{3}%
):0<\ell\leq\sqrt{T}\leq1\}
\]
with its Borel $\sigma$-field $\mathcal{B}\left(  \Xi\right)  $. Let $\left(
\Omega,\mathcal{A},P\right)  $ be a probability space, with expectation
denoted by $E$, and let $\mu_{\omega}$, $\omega\in\Omega$, be a Poisson random
measure on $\mathcal{B}\left(  \Xi\right)  $, with intensity $\nu$ (a $\sigma
$-finite measure on $\mathcal{B}\left(  \Xi\right)  $) given by
\[
d\nu(U,\ell,T,X)=d\gamma(U,\ell,T)d\mathcal{W}(X).
\]
for $\gamma$ a $\sigma$-finite measure on the Borel sets of $\{(U,\ell
,T)\in\mathbb{R}_{+}^{3}:0<\ell\leq\sqrt{T}\leq1\}$ and $d\mathcal{W}(X)$ the
$\sigma$-finite measure defined by
\[
\int_{C([0,1];\mathbb{R}^{3})} \psi(X) d\mathcal{W}(X) = \int_{\mathbb{R}^{3}}
\left[  \int_{C([0,1];\mathbb{R}^{3})} \psi(X) d\mathcal{W}_{x_{0}}(X)\right]
dx_{0}%
\]
for any integrable test function $\psi: C([0,1];\mathbb{R}^{3}) \to\mathbb{R}
$. Here $d\mathcal{W}_{x_{0}}(X)$ is the Wiener measure on $C([0,1],\mathbb{R}%
^{3})$ starting at $x_{0}$ and $dx_{0}$ is the Lebesgue measure on
$\mathbb{R}^{3}$. Heuristically the measure $\mathcal{W}$ describes a Brownian
path starting from an uniformly distributed point in all space. The
assumptions on $\gamma$ will be specified at due time.

The random measure $\mu_{\omega}$ is uniquely determined by its characteristic
function
\[
\mathbb{E}\exp\left(  \int_{\Xi}\varphi(\zeta)\mu_\omega(d\zeta)\right)
=\exp\left(  \int_{\Xi}(e^{\varphi(\zeta)}-1)\nu(d\zeta)\right)\;,
\]
for any bounded measurable function $\varphi$ on $\Xi$ with support in a set
of finite $\nu$-measure. In particular, for example, the first two moments of
$\mu$ read
\[
\mathbb{E}\int_{\Xi}\varphi(\xi)\mu(d\xi)=\int_{\Xi}\varphi(\xi)\nu(d\xi)
\]
and
\[
\mathbb{E}\left[  \int_{\Xi}\varphi(\xi)\mu(d\xi)\right]  ^{2}=\left[
\int_{\Xi}\varphi(\xi)\nu(d\xi)\right]  ^{2}+\int_{\Xi}\varphi^{2}(\xi
)\nu(d\xi).
\]

Given the Poisson random measure $\mu$ we can introduce our random velocity
field as
\begin{equation}
u(x)=\int_{\Xi}u_{\text{single}}^{\zeta}\left(  x\right)  \mu(d\zeta
)=\mu\left(  u_{\text{single}}^{\cdot}\left(  x\right)  \right)  .
\end{equation}
for any $x \in\mathbb{R}^{3}$, where $\zeta= (U,\ell,T,X)$ and
\[
\zeta\mapsto u_{\text{single}}^{\zeta}\left(  x\right)  :=\frac{U}{\ell^{2}%
}\int_{0}^{T}\mathcal{K}_{\ell}(x-X_{t})\wedge dX_{t}%
\]
can be shown to be a well defined $\mu$-measurable function.

In plain words, given $\omega\in\Omega$, the point measure $\mu_{\omega}$
specifies the parameters and locations of infinitely many filaments: formally
\begin{equation}
\mu=\sum_{\alpha\in\mathbb{N}}\delta_{\zeta^{\alpha}}\label{muformal}%
\end{equation}
for a sequence of i.i.d. random points $\left\{  \zeta^{\alpha}\right\}  $
distributed in $\Xi$ according to $\nu$ (this fact is not rigorous since $\nu$
is only $\sigma$-finite, but can be justified by a localisation procedure).
Since the total velocity at a given point $x\in\mathbb{R}^{3}$ should be the
sum of the contributions from each single filament, i.e. in heuristic terms
\begin{equation}
u(x)=\sum_{\alpha}u_{\text{single}}^{\zeta^{\alpha}}\left(  x\right)
\label{uformal}%
\end{equation}
this justifies, physically, the above formula.

To end the construction of the model it remains to choose a suitable measure
$\gamma$ for the distribution of the parameters. Lacking physically motivated
choices of $\gamma$ we resorted in~\cite{FlaGub} to show that it is possible
to fix $\gamma$ in such a way to recover statistics which corresponds to
multifractal scaling of the velocity increments, for any possible choice of
the multifractal spectrum. In this way we showed how to build a random field
with prescribed multifractal spectrum which also possess some geometric
properties of real turbulent fields. In particular we can choose $\gamma$ to
recover K41 behaviour of the velocity increments. In the following we will fix
this particular choice and show that, for our random field
\begin{equation}\label{eq:vortx-scal}
\mathbb{E} |Du(0)|^{2} \sim\eta^{-4/3} \qquad\mathbb{E} |D^{2} u(0)|^{2}
\sim\eta^{-10/3}%
\end{equation}
where $\eta$, in the context of this section will be a UV cutoff scale for the
vortex model, i.e. we will not allow vortices with thickness $\ell$ smaller
than $\eta$ which physically models the ``viscous'' (or Kolmogorov) scale
which determines the lower end of the inertial range. The
scaling~(\ref{eq:vortx-scal}) implies that $\theta\sim\eta$ for $\eta\to0$.

So we stipulate that
\begin{equation}
d\gamma(U,\ell,T)= \delta_{\ell^{1/3}}(U)\delta_{\ell^{2}}(T)\ell^{-4}
1_{\ell \in (\eta,1)} d\ell
\end{equation}
where we ignore a possible constant prefactor which will not play any role in
our discussion. This choice of $\gamma$ corresponds to force the vortex
filaments with thickness $\ell$ to have length proportional to $\ell^{2}$ and
to have typical velocity of the order of $\ell^{1/3}$, the ``density''
$\ell^{-4}$ is chosen to roughly have ``space-filling'' vortices at all scales.
Moreover vortices can have thickness going from the small scale $\eta$ to a large ``integral'' scale of order $1$.

Let us state the result. For technical reasons we will assume that there
exists positive constants $c,C,\lambda$ and $u \in B(0,1)$ such that the
following bounds on $\mathcal{K}$ holds
\begin{equation}
\label{eq:K-bounds-1}
c 1_{x \in B(u,\lambda)} \le|D \mathcal{K}_{1}(x)| \le C 1_{x \in B(0,1)}%
\end{equation}
and
\begin{equation}
\label{eq:K-bounds-1bis}
c 1_{x \in B(u,\lambda)} \le|D^{2} \mathcal{K}_{1}(x)| \le C 1_{x \in B(0,1)} .
\end{equation}
for any $x \in B(0,1)$. A sufficient condition for the upper-bounds is that
$\rho$ is bounded.

\begin{proposition}
With the above definitions, we have
\begin{equation}
\mathbb{E}|Du(0)|^{2}\asymp\eta^{-4/3}\qquad\mathbb{E}|D^{2}u(0)|^{2}%
\asymp\eta^{-10/3}\label{eq:vortx-cnj}%
\end{equation}
as $\eta\rightarrow0$.
\end{proposition}

\begin{proof}
By a small abuse of notation, we have
\[
\mathbb{E}|Du(0)|^{2}=\nu\left[  |Du_{\text{single}}(0)|^{2}\right]
=\nu\left[  \frac{U^{2}}{\ell^{4}}\int_{0}^{T}|D\mathcal{K}_{\ell}(0-X_{t}%
)|^{2}dt\right]
\]
since $\nu(Du_{\text{single}}(0))=0$ being $u_{\text{single}}$ a $\nu$-It\^{o}
integral and where we used the energy-identity for the It\^{o} integral. Note
that $|D\mathcal{K}_{\ell}(x)|=|D\mathcal{K}_{1}(x/\ell)|$ and that, when $x\not \in B(0,1)$ we
have
\begin{equation}
|D\mathcal{K}_{1}(x)|\leq C|x|^{-2},\qquad|D^{2}\mathcal{K}_{1}(x)|\leq C|x|^{-3}%
\label{eq:K-bounds-3}%
\end{equation}
by direct estimation from the Biot-Savart formula.

Now we use the Lemma~\ref{lemma:wiener-bounds} below together with the
bounds~(\ref{eq:K-bounds-1}) and~(\ref{eq:K-bounds-3}) to get
\[
\mathbb{E}|\xi(0)|^{2}\asymp\gamma\left[  U^{2}\ell^{-1}T\right]
=\gamma\left[  \ell^{5/3}\right]
\]
where we used the fact that, under $\gamma$, $U=\ell^{1/3}$ and $T=\ell^{2}$.
Then easily we conclude that
\[
\mathbb{E}|Du(0)|^{2}\asymp\eta^{-4/3}.
\]
Analogously, from the bounds~(\ref{eq:K-bounds-1bis}) and~(\ref{eq:K-bounds-3})
we have that
\[%
\begin{split}
c\ell^{-1}1_{x\in B(u,\lambda\ell)} &  \leq|D^{2}\mathcal{K}_{\ell}(x)|\leq C\ell
^{-1}\left[  1_{x\in B(0,\ell)}+1_{x\not \in B(0,\ell)}|x/\ell|^{-3}\right] \\
&  \leq C^{\prime}\ell^{-1}\left[  1_{x\in B(0,\ell)}+1_{x\not \in B(0,\ell
)}|x/\ell|^{-2}\right]
\end{split}
\]
so using again Lemma~\ref{lemma:wiener-bounds}, we can obtain that
\[
\mathbb{E}|D^{2}u(0)|^{2}\asymp\gamma\left[  U^{2}\ell^{-3}T\right]
=\gamma\left[  \ell^{-1/3}\right]  \asymp\eta^{-10/3}%
\]
ending the proof.
\end{proof}

\begin{lemma}
\label{lemma:wiener-bounds} We have the estimate
\begin{equation}
\mathcal{W}\left[  \int_{0}^{T}1_{X_{t}\in B(0,\ell)}dt\right]  \asymp\ell
^{3}T\label{eq:asymp-bound-1}%
\end{equation}
and, if
\begin{equation}
|\varphi_{\ell}(x)|\leq C(1_{x\in B(0,\ell)}+1_{x\not \in B(0,\ell)}%
|x/\ell|^{-2})\label{eq:varphi-bound}%
\end{equation}
we have
\begin{equation}
\mathcal{W}\left[  \int_{0}^{T}|\varphi_{\ell}(x-X_{t})|^{2}dt\right]
\leq\ell^{3}T\label{eq:asym-bound-2}%
\end{equation}
\end{lemma}

\begin{proof}
These results are particular cases of more general bounds proved
in~\cite{FlaGub}: the first is proved in Lemma~14 of the reference, while
eq.(\ref{eq:asym-bound-2}) is proved in Lemma~3: the proof refers to the
particular case in which $\varphi_{\ell}=\mathcal{K}_{\ell}$ but it is easy to see that
a sufficient condition is given by eq.(\ref{eq:varphi-bound}).
\end{proof}

\appendix
\section{Mollification of measures}

Some computations of the paper with Taylor formula require more regularity
than that of typical fields under $\mu\in\mathcal{P}$. For this reason we
introduce mollifications of measures $\mu\in\mathcal{P}$. Let us remark that
this technical effort is useless if the noise is more regular, since one can
prove more regularity of the typical elements under $\mu\in\mathcal{P}$.

Let $\varphi:\mathbb{R}\rightarrow\mathbb{R}$ be a smooth function with
compact support, symmetric, non negative, strictly positive at zero, with
$\int_{\mathbb{R}^{d}}\varphi\left(  \left\|  x\right\|  \right)  dx=1$. Set
$\phi_{\varepsilon}\left(  x\right)  = \varepsilon^{-d}\varphi\left(
\left\|  x/\varepsilon\right\|  \right)  $, so $\int_{\mathbb{R}^{d}%
}\phi_{\varepsilon}\left(  x\right)  dx=1$;\ $\left\{  \phi_{\varepsilon
}\right\}  _{\varepsilon>0}$ is a family of usual smooth mollifiers. For every
$u\in H$ set
\[
u_{\varepsilon}\left(  x\right)  =\int_{\mathbb{R}^{d}}\phi_{\varepsilon
}\left(  x-y\right)  u\left(  y\right)  dy.
\]
Given $\mu\in\mathcal{P}_{0}$, the mapping $u\mapsto u_{\varepsilon}$ in $H$
induces an image measure $\mu_{\varepsilon}\in\mathcal{P}_{0}$ which is in
fact supported on smooth fields.

\begin{lemma}
If $\mu\in\mathcal{P}$, then $\mu_{\varepsilon}\in\mathcal{P}$.
\end{lemma}

\begin{proof}
We have
\begin{align*}
u_{\varepsilon}\left(  x-a\right)   &  =\int_{\mathbb{R}^{d}}\phi
_{\varepsilon}\left(  x-a-y\right)  u\left(  y\right)  dy\overset{y^{\prime
}=y+a}{=}\int_{\mathbb{R}^{d}}\phi_{\varepsilon}\left(  x-y^{\prime}\right)
u\left(  y^{\prime}-a\right)  dy^{\prime}\\
&  \overset{\mathcal{L}}{=}\int_{\mathbb{R}^{d}}\phi_{\varepsilon}\left(
x-y^{\prime}\right)  u\left(  y^{\prime}\right)  dy^{\prime}%
\end{align*}
where the last equality is understood in law under $\mu$, and it holds true as
processes in $x$. Hence $u_{\varepsilon}\left(  \cdot-a\right)  \overset
{\mathcal{L}}{=}u_{\varepsilon}\left(  \cdot\right)  $. This means
\[
\int_{H}f\left(  u_{\varepsilon}\left(  \cdot-a\right)  \right)  d\mu\left(
u\right)  =\int_{H}f\left(  u_{\varepsilon}\right)  d\mu\left(  u\right)
\]
for bounded continuous $f$'s, and therefore
\[
\int_{H}f\left(  u\left(  \cdot-a\right)  \right)  d\mu_{\varepsilon}\left(
u\right)  =\int_{H}f\left(  u\right)  d\mu_{\varepsilon}\left(  u\right)
\]
so the space homogeneity of $\mu_{\varepsilon}$ is proved.

Similarly, we have
\[
u_{\varepsilon}\left(  Rx\right)  =\int_{\mathbb{R}^{d}}\phi_{\varepsilon
}\left(  R\left(  x-R^{-1}y\right)  \right)  u\left(  y\right)  dy=\int
_{\mathbb{R}^{d}}\phi_{\varepsilon}\left(  x-R^{-1}y\right)  u\left(
y\right)  dy
\]
form the symmetry of $\phi_{\varepsilon}$, hence
\begin{align*}
&  u_{\varepsilon}\left(  Rx\right)  \overset{y^{\prime}=R^{-1}y}{=}%
\int_{\mathbb{R}^{d}}\phi_{\varepsilon}\left(  x-y^{\prime}\right)  u\left(
Ry^{\prime}\right)  dy^{\prime}\\
&  \overset{\mathcal{L}}{=}\int_{\mathbb{R}^{d}}\phi_{\varepsilon}\left(
x-y^{\prime}\right)  Ru\left(  y^{\prime}\right)  dy^{\prime}%
\end{align*}
hence
\[
\int_{H}f\left(  u_{\varepsilon}\left(  R\cdot\right)  \right)  d\mu\left(
u\right)  =\int_{H}f\left(  Ru_{\varepsilon}(\cdot)\right)  d\mu\left(
u\right)
\]
and finally
\[
\int_{H}f\left(  u\left(  R\cdot\right)  \right)  d\mu_{\varepsilon}\left(
u\right)  =\int_{H}f\left(  Ru(\cdot)\right)  d\mu_{\varepsilon}\left(
u\right)  .
\]
The proof is complete.
\end{proof}

\begin{lemma}
For every $\mu\in\mathcal{P}$, if
\[
\int_{H}\int_{\mathcal{T}}\left\|  Du\left(  x\right)  \right\|  ^{2}%
dx\,d\mu\left(  u\right)  <\infty,\quad\int_{H}\int_{\mathcal{T}}\left\|
D^{2}u\left(  x\right)  \right\|  ^{2}dx\,d\mu\left(  u\right)  <\infty,
\]
then
\[
\int_{H}\left\|  u\left(  re\right)  -u\left(  0\right)  \right\|  ^{2}%
d\mu\left(  u\right)  <\infty
\]
and
\[
\lim_{\varepsilon\rightarrow0}\int_{H}\left\|  Du\left(  0\right)  \right\|
^{2}d\mu_{\varepsilon}\left(  u\right)  =\int_{H}\int_{\mathcal{T}}\left\|
Du\left(  x\right)  \right\|  ^{2}dx\,d\mu\left(  u\right)
\]%
\[
\lim_{\varepsilon\rightarrow0}\int_{H}\left\|  D^{2}u\left(  0\right)
\right\|  ^{2}d\mu_{\varepsilon}\left(  u\right)  =\int_{H}\int_{\mathcal{T}%
}\left\|  D^{2}u\left(  x\right)  \right\|  ^{2}dx\,d\mu\left(  u\right)
\]%
\[
\lim_{\varepsilon\rightarrow0}\int_{H}\left\|  u\left(  re\right)  -u\left(
0\right)  \right\|  ^{2}d\mu_{\varepsilon}\left(  u\right)  =\int_{H}\left\|
u\left(  re\right)  -u\left(  0\right)  \right\|  ^{2}d\mu\left(  u\right)  .
\]
\end{lemma}

\begin{proof}
There exists $C>0$ such that
\[
\int_{\mathcal{T}}\left\|  Du_{\varepsilon}\left(  x\right)  \right\|
^{2}dx\leq C\int_{\mathcal{T}}\left\|  Du\left(  x\right)  \right\|  ^{2}dx
\]
for every $u\in D(A)$; and $\int_{\mathcal{T}}\left\|  Du_{\varepsilon}\left(
x\right)  \right\|  ^{2}dx\rightarrow\int_{\mathcal{T}}\left\|  Du\left(
x\right)  \right\|  ^{2}dx$ as $\varepsilon\rightarrow0$ for every $u\in
D(A)$. Hence, by Lebesgue theorem,
\[
\lim_{\varepsilon\rightarrow0}\int_{H}\left[  \int_{\mathcal{T}}\left\|
Du\left(  x\right)  \right\|  ^{2}dx\right]  d\mu_{\varepsilon}\left(
u\right)  =\int_{H}\int_{\mathcal{T}}\left\|  Du\left(  x\right)  \right\|
^{2}dx\,d\mu\left(  u\right)  .
\]
But $\mu_{\varepsilon}$ is space homogeneous, hence
\[
\int_{H}\left[  \int_{\mathcal{T}}\left\|  Du\left(  x\right)  \right\|
^{2}dx\right]  d\mu_{\varepsilon}\left(  u\right)  =\int_{H}\left\|  Du\left(
0\right)  \right\|  ^{2}d\mu_{\varepsilon}\left(  u\right)  .
\]
This proves the first claim. The proof of the second one is entirely similar.
For the third one, we have
\begin{align*}
\left\|  u_{\varepsilon}\left(  x+re\right)  -u_{\varepsilon}\left(  x\right)
\right\|  ^{2} &  =\left\|  r\int_{0}^{1}Du_{\varepsilon}\left(  x+\sigma
e\right)  ed\sigma\right\|  ^{2}\\
&  \leq r^{2}\int_{0}^{1}\left\|  Du_{\varepsilon}\left(  x+\sigma e\right)
\right\|  ^{2}d\sigma
\end{align*}
for every $u\in D(A)$, hence
\begin{align*}
\int_{\mathcal{T}}\left\|  u_{\varepsilon}\left(  x+re\right)  -u_{\varepsilon
}\left(  x\right)  \right\|  ^{2}dx &  \leq r^{2}\int_{0}^{1}\int
_{\mathcal{T}}\left\|  Du_{\varepsilon}\left(  x+\sigma e\right)  \right\|
^{2}dx\,d\sigma\\
&  =r^{2}\int_{0}^{1}\int_{\mathcal{T}}\left\|  Du_{\varepsilon}\left(
x\right)  \right\|  ^{2}dx\,d\sigma\\
&  \leq Cr^{2}\int_{\mathcal{T}}\left\|  Du\left(  x\right)  \right\|  ^{2}dx.
\end{align*}
Therefore, again by Lebesgue theorem,
\begin{align*}
&  \lim_{\varepsilon\rightarrow0}\int_{H}\int_{\mathcal{T}}\left\|  u\left(
x+re\right)  -u\left(  x\right)  \right\|  ^{2}dx\,d\mu_{\varepsilon}\left(
u\right) \\
&  =\int_{H}\int_{\mathcal{T}}\left\|  u\left(  x+re\right)  -u\left(
x\right)  \right\|  ^{2}dx\,d\mu\left(  u\right)\; .
\end{align*}
The third claim follows now from the space homogeneity of both $\mu
_{\varepsilon}$ and $\mu$.
\end{proof}

We are now in the position to prove a quantitative consequence of isotropy,
that we shall use in the sequel. In the next statement we understand that both
terms in the equality are either finite and equal, or both infinite.

\begin{lemma}
\label{lemmaisotropy}For every $\mu\in\mathcal{P}$ and every coordinate
unitary vector $e$ we have
\[
\int_{H}\int_{\mathcal{T}}\left\|  Du\left(  x\right)  \right\|  ^{2}%
dx\,d\mu\left(  u\right)  =d\int_{H}\int_{\mathcal{T}}\left\|  Du\left(
x\right)  \cdot e\right\|  ^{2}dx\,d\mu\left(  u\right)  .
\]
For $\mu_{\varepsilon}$, we have the same identity and also
\[
\int_{H}\left\|  Du\left(  0\right)  \right\|  ^{2}d\mu_{\varepsilon}\left(
u\right)  =d\int_{H}\left\|  Du\left(  0\right)  \cdot e\right\|  ^{2}%
d\mu_{\varepsilon}\left(  u\right)  .
\]
\end{lemma}

\begin{proof}
\textbf{Step 1}. Denote by $e_{1},...,e_{d}$ the coordinate unitary vectors.
For $u\in D(A)$ we have
\[
\left\|  Du\left(  x\right)  \right\|  ^{2}=\sum_{ij}\left|  \frac{\partial
u_{i}}{\partial x_{j}}\left(  x\right)  \right|  ^{2},\quad\left\|  Du\left(
x\right)  \cdot e_{j}\right\|  ^{2}=\sum_{i}\left|  \frac{\partial u_{i}%
}{\partial x_{j}}\left(  x\right)  \right|  ^{2}%
\]
and thus
\[
\left\|  Du\left(  x\right)  \right\|  ^{2}=\sum_{j}\left\|  Du\left(
x\right)  \cdot e_{j}\right\|  ^{2}.
\]
Therefore
\[
\int_{H}\left\|  Du\left(  0\right)  \right\|  ^{2}d\mu_{\varepsilon}\left(
u\right)  =\sum_{j}\int_{H}\left\|  Du\left(  0\right)  \cdot e_{j}\right\|
^{2}d\mu_{\varepsilon}\left(  u\right)
\]
and
\[
\int_{H}\int_{\mathcal{T}}\left\|  Du\left(  x\right)  \right\|  ^{2}%
dx\,d\mu\left(  u\right)  =\sum_{j}\int_{H}\int_{\mathcal{T}}\left\|  Du\left(
x\right)  \cdot e_{j}\right\|  ^{2}dx\,d\mu\left(  u\right)  .
\]
It is then sufficient to prove that all terms of the sums on the
right-hand-sides are equal, in order to prove the first and last claim of the
lemma; we shall prove this below in steps 2 and 3. Finally, the first
assertion for $\mu_{\varepsilon}$ is a particular case of the first claim of
the lemma ($\mu_{\varepsilon}$ is an element of $\mathcal{P}$).

\textbf{Step 2}. Now, given $j=1,...,d$, take a rotation $R$ as in the
definition of $\mathcal{P}$ such that $Re_{1}=e_{j}$. Given $N>0$,
\begin{align*}
&  \int_{H}\left(  \left\|  Du\left(  0\right)  \cdot e_{j}\right\|
^{2}\wedge N\right)  d\mu_{\varepsilon}\left(  u\right) \\
&  =\lim_{r\rightarrow0}\int_{H}\left(  r^{-2}\left\|  u\left(  re_{j}\right)
-u\left(  0\right)  \right\|  ^{2}\wedge N\right)  d\mu_{\varepsilon}\left(
u\right) \\
&  =\lim_{r\rightarrow0}\int_{H}\left(  r^{-2}\left\|  u\left(  Rre_{1}%
\right)  -u\left(  R0\right)  \right\|  ^{2}\wedge N\right)  d\mu
_{\varepsilon}\left(  u\right) \\
&  =\lim_{r\rightarrow0}\int_{H}\left(  r^{-2}\left\|  u\left(  re_{1}\right)
-u\left(  0\right)  \right\|  ^{2}\wedge N\right)  d\mu_{\varepsilon}\left(
u\right) \\
&  =\int_{H}\left(  \left\|  Du\left(  0\right)  \cdot e_{1}\right\|
^{2}\wedge N\right)  d\mu_{\varepsilon}\left(  u\right)  .
\end{align*}
By monotone convergence in $N$, we get that $\int_{H}\left\|  Du\left(
0\right)  \cdot e_{j}\right\|  ^{2}d\mu_{\varepsilon}\left(  u\right)  $ is
independent of $j$. This proves one of the claims.

\textbf{Step 3}. From the previous step and homogeneity we have that $\int
_{H}\int_{\mathcal{T}}\left\|  Du\left(  x\right)  \cdot e_{j}\right\|
^{2}dx\,d\mu_{\varepsilon}\left(  u\right)  $ is also independent of $j$.
Arguing as in the proof of the previous lemma, this integral converges to
$\int_{H}\int_{\mathcal{T}}\left\|  Du\left(  x\right)  \cdot e_{j}\right\|
^{2}dx\,d\mu\left(  u\right)  $, which is therefore also independent of $j$. The
proof is complete.
\end{proof}

\section{Scaling theorems}\label{appendixtwo}

The torus, $\mathcal{T}_{L}=\left[  0,L\right]  ^{d}$, the energy space
$H_{L}$ with norm $\left|  .\right|  _{H_{L}}$, the spaces $V_{L}$, $D\left(
A_{L}\right)  $, $\mathcal{D}_{L}$ and the Stokes operator $A_{L}$ on $T_{L}$
have been already introduced in section \ref{section 1.1}. We define
\begin{equation}\label{LambdaL}
\Lambda^{(\infty)}_L=\left\{ k\in\frac{2\pi}{L}\ZZ^d\ :\ |k|^2>0\ \right\},
\end{equation}
and, for the purpose of Galerkin approximations, we introduce also
\[
\Lambda^{(n)}_L
=\left\{ k\in\frac{2\pi}{L}\ZZ^d\ :\ 0<|k|^2\leq\left(\frac{2\pi}{L}n\right)^2\right\}
\]
so that $\Lambda^{(\infty)}_L=\cup_{n}\Lambda^{(n)}_L$. In particular,
$\Lambda^{(\infty)}=\Lambda^{(\infty)}_1$. 

\subsection{Scaling theorem for Galerkin approximations}

Let $V_{L}^{\prime}$ be the dual of $V_{L}$; with proper identifications we
have $V_{L}\subset H_{L}\subset V_{L}^{\prime}$ with continuous injections.
Let $B_{L}\left(  .,.\right)  :V_{L}\times V_{L}\rightarrow V_{L}^{\prime}$ be
the bilinear operator defined for all $u,v,w\in\mathcal{D}_{L}$ as
\begin{equation}\label{nonlinop}
\left\langle w,B_{L}\left(  u,v\right)  \right\rangle _{H_{L}}=\sum
_{i,j=1}^{d}\frac{1}{L^d}\int_{\mathcal{T}_L}u_{i}\frac{\partial v_{j}}{\partial x_{i}}%
w_{j}dx
=\sum_{h+l=k}\left(l\cdot\widehat{u}(h)\right)\widehat{v}(l)\cdot
\overline{\widehat{w}(k)}.
\end{equation}

Given $L>0$, $\nu>0$ and $\theta>0$, consider (formally) the equation in $H_{L}$
\[
du+\left[  \nu A_{L}u+B_{L}\left(  u,u\right)  \right]  dt=\theta\sum_{
k\in\Lambda^{(\infty)}_L}\sigma_k^L\,d\beta_k^L\,e^{-i k\cdot x},
\]
where $\beta_k^L=\beta_{Lk}$ and $\sigma_k^L=\sigma_{Lk}$, and
$(\beta_k)_{k\in\Lambda^{(\infty)}}$ and $(\sigma_k)_{k\in\Lambda^{(\infty)}}$
have been introduced in Section \ref{subsubsectnoise} and
are subject to the assumptions imposed therein, so that the random fields
\[
W_{L}^{(n)}(t,x)=\sum_{k\in\Lambda^{(n)}_L}\sigma_k^L\beta_k^L(t)
\,e^{-i k\cdot x}
\]
and the field $W_{L}^{\left(  \infty\right)  }\left(  t,x\right)  $ similarly
defined, are space-homogeneous and partially (in the sense of the rotations of
the torus) isotropic.

Let $H_L^{(n)}$ be the subspace of $H_L$ correspondings to the modes
with wavelengths in $\Lambda^{(n)}_L$
and consider the equation in $H_{L}^{\left(  n\right)  }$
\begin{equation}
du^{\left(  n\right)  }+\left[  \nu A_{L}u^{\left(  n\right)  }+\pi
_{L}^{\left(  n\right)  }B_{L}\left(  u^{\left(  n\right)  },u^{\left(
n\right)  }\right)  \right]  dt=\theta\sum_{k
\in\Lambda^{(n)}_L}\sigma_k^L\,d\beta_k^L\,e^{-i k\cdot x}\label{galerkin}%
\end{equation}
where $\pi_{L}^{\left(  n\right)  }$ is the orthogonal projection of $H_{L}$
onto $H_{L}^{\left(  n\right)  }$.

\begin{lemma}
\label{appendixlemmascaling}If $u^{\left(  n\right)  }$ is a solution in
$H_{L} $ of (\ref{galerkin}), with initial condition $u^{\left(  n\right)
}\left(  0\right)  $ and parameters $\left(  \nu,L,\theta\right)  $, then
\[
\widetilde{u}^{\left(  n\right)  }(t,x):=\lambda^{\beta}u^{\left(  n\right)
}(\lambda^{1+\beta}t,\lambda x)
\]
is a solution in $H_{L/\lambda}$ of equation (\ref{galerkin}) with initial
condition $\widetilde{u}^{\left(  n\right)  }\left(  0\right)  $ and
parameters $\left(
\nu\lambda^{\beta-1},L/\lambda,\lambda^{\frac{1+3\beta}{2}}\theta\right)  $
(but with new Brownian motions).
\end{lemma}

\begin{proof}
This statement is not clear a priori, especially because of the scaling
transformation of the nonlinear term, so we give all the details.
The solution $u^{(n)}$, as a Fourier series, is given by
\[
u^{(n)}(t,x)=\sum_{k\in\Lambda^{(n)}_L}\widehat{u}^{(n)}(t,k)
\,e^{-i k\cdot x},
\]
and the solution $\widetilde{u}^{(n)}$, as a process in $H_{L/\lambda}^{(n)}$,
is given by
\[
\widetilde{u}^{(n)}(t,x)=\sum_{k\in\Lambda^{(n)}_{{L/\lambda}}}
\widehat{\widetilde{u}}^{(n)}(t,k)\,e^{-i k\cdot x},
\]
The Fourier coefficients of $u^{(n)}$ and $\widetilde{u}^{(n)}$ are related
by the scaling
\begin{align}\label{fourierscaling}
\widehat{\widetilde{u}}^{(n)}(t,k)
&=\frac{\lambda^d}{L^d}\int_{\mathcal{T}_{{L/\lambda}}}
  \widetilde{u}^{(n)}(t,x)\,e^{ik\cdot x}\,dx
 =\frac{\lambda^{d+\beta}}{L^d}\int_{\mathcal{T}_{{L/\lambda}}}
  u^{(n)}(\lambda^{1+\beta}t,\lambda x)\,e^{ik\cdot x}\,dx\\
&\overset{\left(\substack{\mbox{\tiny$x'=\lambda x$}\\\mbox{\tiny$k'=k/\lambda$}}\right)}{=}
  \frac{\lambda^\beta}{L^d}\int_{\mathcal{T}_L}
  u^{(n)}(\lambda^{1+\beta}t,x')\,e^{ik'\cdot x'}\,dx'
 =\lambda^\beta\widehat{u}(\lambda^{1+\beta}t,k').\notag
\end{align}
From the equation (\ref{galerkin}) in integral form,
\begin{align*}
& u^{\left(  n\right)  }(t)+\int_{0}^{t}\left[  \nu A_{L}u^{\left(  n\right)
}+\pi_{L}^{\left(  n\right)  }B_{L}\left(  u^{\left(  n\right)  },u^{\left(
n\right)  }\right)  \right]  \left(  s\right)\,ds\\
& =u^{\left(  n\right)  }(0)+\theta\sum_{k\in\Lambda^{(n)}_L}\sigma_k^L
\beta_k^L(t)\,e^{-ik\cdot x},
\end{align*}
we have
\begin{align*}
& \lambda^{\beta}u^{\left(  n\right)  }(\lambda^{1+\beta}t,\lambda x)\\
& +\lambda^{1+2\beta}\int_{0}^{t}\left[  \nu A_{L}u^{\left(  n\right)  }%
+\pi_{L}^{\left(  n\right)  }B_{L}\left(  u^{\left(  n\right)  },u^{\left(
n\right)  }\right)  \right]  \left(  \lambda^{1+\beta}s,\lambda x\right)\,ds=\\
& =\lambda^{\beta}u^{\left(  n\right)  }(0,\lambda x)+\lambda^{\frac{1+3\beta
}{2}}\theta\sum_{k\in\Lambda^{(n)}_{{L/\lambda}}}\sigma_k^{{L/\lambda}}
\widetilde{\beta}_k^{{L/\lambda}}(t)
\,e^{-ik\cdot x}
\end{align*}
where $\widetilde{\beta}_k^{{L/\lambda}}(t):=\lambda^{-\frac{1+\beta}{2}}
\beta_{k/\lambda}^L(\lambda^{1+\beta}t)$ are new Brownian motions. 
The first term on the {l.\,h.\,s.} is $\widetilde{u}^{\left(n\right)}(t,x)$,
and the first term on the {r.\,h.\,s.} is $\widetilde{u}^{(n)}(0,x)$.
In addition, we have
\[
A_{L/\lambda}\widetilde{u}^{\left(  n\right)  }(t,x)=\lambda^{2+\beta}\left(
A_{L}u^{\left(  n\right)  }\right)  (\lambda^{1+\beta}t,\lambda x).
\]
The proof of the claim will be complete if we show that
\[
\lambda^{1+2\beta}\left[  \pi_{L}^{\left(  n\right)  }B_{L}\left(
u^{\left(  n\right)  },u^{\left(  n\right)  }\right)  \right]  \left(
\lambda^{1+\beta}t,\lambda x\right) \\
=\left[  \pi_{L/\lambda}^{\left(  n\right)  }B_{L/\lambda}\left(
\widetilde{u}^{\left(  n\right)  },\widetilde{u}^{\left(  n\right)  }\right)
\right]  (t,x).
\]
For every $\varphi\in V_{L/\lambda}$, by using the Fourier expression
(\ref{nonlinop}) of the non-linear term and the scaling of Fourier coefficients
(\ref{fourierscaling}), 
\begin{align*}
  \langle\pi^{(n)}_{{L/\lambda}}&B_{{L/\lambda}}
  (\widetilde{u}^{(n)},\widetilde{u}^{(n)})(t,\cdot),\varphi\rangle_{H_{{L/\lambda}}}
=\langle B_{{L/\lambda}}(\widetilde{u}^{(n)},\widetilde{u}^{(n)})(t,\cdot),
  \pi^{(n)}_{{L/\lambda}}\varphi\rangle_{H_{{L/\lambda}}}\\
&=\sum_{h+l=k}\left(l\cdot\widehat{\widetilde{u}}^{(n)}(t,h)\right)
  \widehat{\widetilde{u}}^{(n)}(t,l)\cdot\overline{\widehat{\varphi}(k)}\\
&=\lambda^{1+2\beta}\sum_{h+l=k}\left(l\cdot\widehat{u}^{(n)}(\lambda^{1+\beta}t,\frac{h}{\lambda})\right)
  \widehat{u}^{(n)}(\lambda^{1+\beta}t,\frac{l}{\lambda})\cdot\overline{\widehat{\varphi}(k)}\\
&=\lambda^{1+2\beta}\langle B_L(u^{(n)},u^{(n)})(\lambda^{1+\beta}t,\lambda\cdot),
  \pi^{(n)}_{{L/\lambda}}\varphi\rangle_{H_{{L/\lambda}}}\\
&=\lambda^{1+2\beta}\langle\pi^{(n)}_LB_L(u^{(n)},u^{(n)})
  (\lambda^{1+\beta}t,\lambda\cdot),\varphi\rangle_{H_{{L/\lambda}}},
\end{align*}
where the sums above are extended to all wavelengths $h$, $l$ and $k\in\Lambda^{(n)}_{{L/\lambda}}$
such that $h+l=k$.
\end{proof}

\subsection{Scaling theorem for stationary measures}

Similarly to section \ref{subsub3D}, denote by $\mathcal{P}_{NS}^{G}\left(
\nu,L,\theta\right)  $ the set of probability measures that are limit of
homogeneous isotropic invariant measures of equations (\ref{galerkin}).

Given $\lambda>0$ and $\beta\in\mathbb{R}$ and $\mu\in\mathcal{P}_{NS}%
^{G}\left(  \nu,L,\theta\right)  $, let $u$ be a random field on $T_{L}$ with
law $\mu$, define the random field $\widetilde{u}$ on $T_{L/\lambda}$ as
\[
\widetilde{u}(x)=\lambda^{\beta}u(\lambda x)
\]
and let $\widetilde{\mu}$ be the law of $\widetilde{u}$ on $H_{L/\lambda}$.
More intrinsically, $\widetilde{\mu}$ is defined by the relation
\[
\int_{H_{L/\lambda}}f(u)\,d\widetilde{\mu}(u)=\int_{H_{L}}f(\lambda^{\beta
}u\left(  \lambda\cdot\right)  )\,d\mu(u)
\]
for every bounded continuous $f$ on $H_{L/\lambda}$.

\begin{theorem}
\label{appendixtheoremscaling}\bigskip If $\mu\in\mathcal{P}_{NS}^{G}\left(
\nu,L,\theta\right)  $ then $\widetilde{\mu}\in\mathcal{P}_{NS}^{G}\left(
\nu\lambda^{\beta-1},L/\lambda,\lambda^{\frac{1+3\beta}{2}}\theta\right)  $.
\end{theorem}

\begin{proof}
The measure $\mu$ of the theorem is the weak limit of a sequence $\left\{
\mu_{n_{k}}\right\}  $ of invariant measures on $H_L^{\left(  n_{k}\right)
} $ of the Galerkin problems with indexes $n_{k}$. For each $n_{k}$, let
$u^{\left(  n_{k}\right)  }$ be a stationary solution (on some probability
space) of (\ref{galerkin}), with parameters $\left(  \nu,L,\theta\right)  $
and marginal $\mu_{n_{k}}$. Let $\widetilde{u}^{\left(  n_{k}\right)  }$ be
the rescaled process as above, which is a solution of (\ref{galerkin}) with
parameters $\left(  \nu\lambda^{\beta-1},L/\lambda,\lambda^{\frac{1+3\beta}%
{2}}\theta\right)  $ (by the lemma above) and is a stationary process. Its
marginal $\widetilde{\mu}_{n_{k}}$ is the scaling of $\mu_{n_{k}}$, similarly
to the relation defined above between $\mu$ and $\widetilde{\mu}$. Moreover
$\widetilde{\mu}_{n_{k}}$ is an invariant measure for equation (\ref{galerkin}%
) with parameters $\left(  \nu\lambda^{\beta-1},L/\lambda,\lambda
^{\frac{1+3\beta}{2}}\theta\right)$. From the weak convergence of
$\mu_{n_{k}}$ to $\mu$ it is now easy to deduce the weak convergence of
$\widetilde{\mu}_{n_{k}}$ to $\widetilde{\mu}$. Therefore $\widetilde{\mu}%
\in\mathcal{P}_{NS}^{G}\left(  \nu\lambda^{\beta-1},L/\lambda,\lambda
^{\frac{1+3\beta}{2}}\theta\right)  $. The proof is complete.
\end{proof}


\def\polhk#1{\setbox0=\hbox{#1}{\ooalign{\hidewidth
  \lower1.5ex\hbox{`}\hidewidth\crcr\unhbox0}}}

\end{document}